\documentclass[]{emulateapj}

\slugcomment{Accepted by The Astrophysical Journal}

\shorttitle{Mergers of Unequal Mass Galaxies}
\shortauthors{F. M. Khan et al.}

\begin{document}

%% LaTeX will automatically break titles if they run longer than
%% one line. However, you may use \\ to force a line break if
%% you desire.

\title{Mergers of Unequal Mass Galaxies: Supermassive Black Hole Binary Evolution and Structure of Merger Remnants}

%% Use \author, \affil, and the \and command to format
%% author and affiliation information.
%% Note that \email has replaced the old \authoremail command
%% from AASTeX v4.0. You can use \email to mark an email address
%% anywhere in the paper, not just in the front matter.
%% As in the title, use \\ to force line breaks.

%\author{Fazeel Mahmood Khan\altaffilmark{1,2}, Miguel Preto\altaffilmark{1}, Peter Berczik\altaffilmark{1,3,4}, Ingo Berentzen\altaffilmark{1}
%Andreas Just\altaffilmark{1}, Rainer Sourzem\altaffilmark{1}}
%\affil{$^1$Astronomisches Rechen-Institut, Zentrum f\"ur Astronomie, 
%Univ. of Heidelberg, M{\"o}nchhof-Strasse 12-14, 69120 Heidelberg, Germany\\
%$^2$Department of Physics, Government College University (GCU), 54000  
%Lahore, Pakistan\\
%$^3$National Astronomical Observatories of China, Chinese Academy of Sciences, 20A Datun 
%Rd., Chaoyang District, 100012, Beijing, China\\
%$^4$Main Astronomical Observatory, National Academy of Sciences of Ukraine, 27 Akademika 
%Zabolotnoho St., 03680, Kyiv, Ukraine\\
%}
\author{Fazeel Mahmood Khan\altaffilmark{1,2}}
\author{Miguel Preto\altaffilmark{1}}
\author{Peter Berczik\altaffilmark{3,1,4}}
\author{Ingo Berentzen\altaffilmark{1,5}}
\author{Andreas Just\altaffilmark{1}}
\author{Rainer Spurzem\altaffilmark{3,1,6}}

\affil{$^1$Astronomisches Rechen-Institut, Zentrum f\"ur Astronomie, 
Univ. of Heidelberg, M{\"o}nchhof-Strasse 12-14, 69120 Heidelberg, Germany}
\affil{$^2$Department of Physics, Government College University (GCU), 54000  
Lahore, Pakistan}
\affil{$^3$National Astronomical Observatories of China, Chinese Academy of Sciences, 20A Datun 
Rd., Chaoyang District, 100012, Beijing, China}
\affil{$^4$Main Astronomical Observatory, National Academy of Sciences of Ukraine, 27 Akademika 
Zabolotnoho St., 03680, Kyiv, Ukraine}
\affil{$^5$Heidelberg Institute for Theoretical Studies, Schlosswolfsbrunnenweg 35, 69118 Heidelberg, Germany}
\affil{$^6$The Kavli Institute for Astronomy and Astrophysics at Peking University (KIAA), Yi He Yuan Lu 5, Hai Dian Qu, Beijing, 100871, China}

\begin{abstract}
%%At a separation of approximately one parsec, binary supermassive black holes (SMBHs) stall in spherial galaxy model. In an earlier %%study, we show that stalling does not happen for equal mass galaxy mergers. Galaxies that form via major mergers are substantially %%nonspherical, and it has been argued that the centrophilic orbits in triaxial galaxies might provide stars to the massive binary at a %%high enough rate to avoid stalling. Here using a set of direct N-body simulations of unequal mass mergers with different galaxy(and %%SMBH) mass ratios for different density profiles of merging galaxies, we study the shape of merger remnat and the SMBH binary %%hardening rates, we find .................
Galaxy centers are residing places for Super Massive Black Holes (SMBHs). Galaxy mergers bring SMBHs close together
to form gravitationally bound binary systems which, if able to coalesce in less than a Hubble time, would be one 
of the most promising sources of gravitational waves for the {\it Laser Interferometer Space Antenna} (LISA). In spherical 
galaxy models, SMBH binaries stall at a separation of approximately one parsec, leading to the ``final parsec problem"
(FPP). On the other hand, it has been shown that merger-induced triaxiality of the remnant in equal-mass mergers is
capable of supporting a 
constant supply of stars on so-called centrophilic orbits that interact with the binary and thus avoid the FPP. In this paper, 
using a set of direct $N$-body simulations of mergers of initially spherically symmetric galaxies with different mass ratios, we 
show that the merger-induced triaxiality is also able to drive unequal-mass SMBH binaries to coalescence. The binary 
hardening rates are high and depend only weakly on the mass ratios of SMBHs for a wide range of mass ratios $q$. There is, 
however, an abrupt transition in the hardening rates for mergers with mass ratios somewhere between $q \sim 0.05$ and 
$0.1$, resulting from the monotonic decrease of merger-induced triaxiality with mass ratio $q$, as the secondary galaxy 
becomes too small and light to significantly perturb the primary, i.e., the more massive one. The hardening rates are
significantly higher for galaxies having steep cusps in comparison with those having shallow cups at centers. The 
evolution of the binary SMBH leads to relatively shallower inner slopes at the centers of the merger remnants. The stellar
mass displaced by the SMBH binary on its way to coalescence is $\sim 1-5$ times the combined mass of binary SMBHs. The
coalescence time scales for SMBH binary with mass $ \sim 10^{6} M_\odot$ are 
less than 1 Gyr and for those at the upper end of SMBH masses $10^{9} M_\odot$ are 1-2 Gyr for less eccentric binaries 
whereas less than 1 Gyr for highly eccentric binaries. SMBH binaries are thus expected to be promising sources of 
gravitational waves at low and high redshifts.

\end{abstract}

%% Keywords should appear after the \end{abstract} command. The uncommented
%% example has been keyed in ApJ style. See the instructions to authors
%% for the journal to which you are submitting your paper to determine
%% what keyword punctuation is appropriate.


\section{Introduction}\label{sec-intro}

Super Massive Black Holes (SMBHs) are now a well established component of galaxies having a sizable bulge \citep{FF05}. In the paradigm
of the $\Lambda$ Cold Dark Matter cosmology, galaxies are formed via hierarchical merging. If more than one of these merging galaxies 
contain a SMBH, the formation of a binary SMBH is almost inevitable \citep{BBR80}. There are cases in which there is a clear 
observational evidence for two widely separated SMBHs as well as some circumstantial evidence for a true SMBH binary \citep{komossa06}. 
Upon coalescence, SMBH binaries would be the highest signal-to-noise ratio sources of gravitational waves, detectable from essentially 
any cosmological redshift, for future space-borne gravitational wave detectors such as the {\it Laser Interferometer Space Antenna} 
(LISA) \citep{Hughes03,BC04,LISA2011}.  
 
The paradigm for SMBH binary evolution, after a merger of gas-poor galaxies, consists of three distinct phases \citep{BBR80}. 
First, the two SMBHs sink towards the center due to the dynamical friction exerted by the stars and the dark matter. Dynamical
friction acts
together with ''gravitational slingshot effect`` as SMBHs form a bound pair with semi-major axis $a \sim r_\mathrm{h}$, where $r_\mathrm{h}$ is 
the binary's influence radius, which typically is defined to be the radius which encloses twice the mass of the binary in 
stars $M(<r_\mathrm{h})=2 M_\bullet$, where $M_\bullet$ is the mass of the binary. Dynamical friction stops being an
effective driver of inspiral when the binary reaches the {\it hard binary} separation $a \sim a_\mathrm{h}$ \citep{Q96,Yu02}
\begin{equation}
a_\mathrm{h} := \frac{G \mu_r}{4 \sigma^2},
\label{eqn1}
\end{equation}
where $\mu_\mathrm{r}$ is the binary's reduced mass and $\sigma$ is the $1$D velocity dispersion. In the second phase, 
{\it slingshot} ejection of stars is the dominant mechanism for taking energy and angular momentum away from 
SMBH binary. Thirdly, the binary eventually reaches a separation $a_\mathrm{GW}$ at which the loss of orbital energy 
due to Gravitational Wave (GW) emission drives the final coalescence. The transition from the first to the 
second phase is prompt provided that the mass ratio of the remnants is not too small $q \gtrsim 0.1$ 
\citep{calleg11}. In contrast, the subsequent
transition from the second to the third phase could constitute a bottleneck for the binary
evolution towards final coalescence caused by lack of stars which can interact with the binary. This is
the so-called {\it Final Parsec Problem}. 

\cite{Yu02} early pointed out, based on an analysis based on the HST sample of nearby elliptical galaxies and spiral 
bulges from \cite{Faber97}, that flattening and non-axisymmetries would help stellar dynamics in bringing SMBH binaries 
to coalescence. \cite{MP2004} built self-consistent cuspy, triaxial models with a single SMBH at the center and showed 
that such models could be concocted with a significant fraction of centrophilic orbits that would efficiently drive 
the hardening rate of a binary if it were present at the center. \cite{ber06} showed, for the first time with $N$-body 
simulations of rotating King models, that the FPP could potentially be solved by purely stellar dynamical models that 
develop a significant amount of triaxiality. By including the post-Newtonian ($\mathcal{PN}$) terms into the
equations of motion of the binary, \cite{berent09} explicitly showed that the coalescence times, if the latter 
models are scaled to real galaxies, are clearly shorter than a Hubble time. \cite{Preto09} presented preliminary 
results suggestive that merger-induced triaxiality resulting from the merger of spherically symmetric models 
would lead to qualitatively similar results -- namely the prompt coalescence of resulting SMBH binaries. 
\cite{kh11} (hereafter paper I) and \cite{Preto11} (hereafter paper II) confirmed and generalized the latter 
results by showing that galaxies which form via equal mass mergers are mildly triaxial. They have shown that hardening 
rates of SMBH binaries resulting from equal-mass mergers are substantially higher than those found in spherical nuclei, 
and are independent of the total number $N$ of stars, thus allowing them to extrapolate results to real galaxies. 
Presumably, the nonspherical shapes of merger remnants reported in these studies result in a large population of stars 
on centrophilic orbits. Inside the influence radius of SMBH, the centrophilic orbit family includes saucer or cone orbits 
in the axisymmetric potentials \citep{st99} and pyramids orbits (angular momentum close to zero) in triaxial potentials 
\citep{mv11}. Outside the influence radius, most of the centrophilic orbits are choatic in a triaxial potential \citep{pm01}.

The inspiral of a binary SMBH is expected to leave a characteristic imprint in the morphological and dynamical 
properties of the newly formed galactic nucleus following the merger: e.g., the bending and precession of radio 
jets \citep{roos93}, the variability in optical light curve of quasar \citep{val08}, X-shaped radio lobes \citep{mer-ek02} 
and, with some likelihood, a partially depleted core instead of a steep cusp in bright elliptical galaxies 
\citep{Gra04,kormendy09} could be explained by different models of SMBH binary evolution prior or after coalescence 
\citep{MMS2007,gm08}. 

Detailed surface photometry of ellipticals has revealed that their surface brightness is well described by a S{\'e}rsic
profile, $\log I(r) \propto r^{1/n}$, over most of the body of a bulge or early-type galaxy. However, systematic 
deviations from this profile are found in the innermost regions close to the central SMBH -- either in the form of a light 
(mass) excess or deficit with respect to that would result from the extrapolation of the S{\'e}rsic profile 
towards the center \citep{kormendy09}. In other words, observations seem to reveal a dichotomy in the central density profiles 
of bulges and early-type galaxies: giant, high-luminosity objects have relatively shallow central density profiles while 
normal, low-luminosity ones show steeper density profiles in their center. The former are often called {\it core} galaxies 
and show density profiles with central logarithmic slopes $\gamma < 1$, and typically harbor SMBH of mass $\gtrsim 10^8 
M_\odot$; while the latter are normally called {\it power law} galaxies, have central $\gamma > 1$ and lighter SMBH
$\lesssim {\rm few} \times 10^7 M_\odot$. The explanation of cores is non-trivial since dry mergers should preserve 
the highest density in phase space given that, by virtue of being devoid of appreciable amounts of gas, their dynamics 
is essentially collisionless \citep{BT08}.

Galaxies having a single mass stellar population are expected to have \citet{BW76} cusp around the SMBH with 
$\rho(r) \propto r^{-\alpha}, \alpha \sim 7/4$ for $r \lesssim  r_\mathrm{h}$ after a relaxtion time \citep{Preto04}. In the more realistic case where there is a
range of stellar masses, less massive objects follow a shallow profile, $\alpha \sim 3/2$, while heavier objects (e.g. stellar black
holes) develop steeper cusps $\alpha \sim 2$ around the SMBH \citep{BW77,ale09}. Whether
a given, or the typical, nucleus has reached this relaxed state or not
depends on its ``initial" profile and on how long ago it was formed. Different studies have reached different
conclusions regarding the most probable state of the stellar distribution
around the SMBH after a relaxation time. \citet{PA10}, who performed
Fokker-Planck and $N-$body simulations of a two component galaxy nucleus model made of solar mass stars and stellar mass black
holes, obtained mass-segregated stellar cusps in $\sim$ 1/4 of a relaxation time.
%; whereas\cite{gua12} who studied the regrowth of the cusp around SMBH after the merger of galaxies containing collisionally-relaxed nuclei of four species (solar mass stars, white %dwarfs, neutron stars and 10 M$\odot$ black holes) did not find such state after a few relaxation times. 
However, \citet{gua12} state roughly the opposite conclusion: that even after a few central relaxation times, the distribution of stars can be very different than in the steady-state, mass-segregated solutions.

The analysis of the number counts of spectroscopically identified, old stars in the sub-parsec region of our own Milky Way \citep{buch09,do09} seems to favor the presence of a very shallow profile for the {\it visible} fraction of the old stellar population. This is consistent with the finding that the two-body relaxation time at the influence radius 
of the SMBH in the center of the Milky Way is longer than a Hubble time \citep{mer10,PA10}. On the other hand, low-luminosity spheroids often contain nuclear star clusters which have 
central relaxation times shorter than a Hubble time \citep{mer09}.

In dry mergers,  i.e. in the absence of a dynamically significant amount of gas, the slingshot ejection of stars 
by the central SMBH binary drives the binary inspiral. It has early been proposed that inspiraling binaries could carve 
a core inside $\sim r_\mathrm{h}$, thus providing an explanation for the mass deficits present in the center of gas-poor giant ellipticals 
\citep{ME96}. On energetic grounds it can be shown that in order for the binary to reach coalescence, it needs to eject an 
amount of stellar mass of order of its own mass, $M_\mathrm{ej} \sim \alpha \times M_{\bullet}$, where $\alpha = O(1)$ and $M_{\bullet} = M_{\bullet 1} 
+ M_{\bullet 2}$ \citep{LR2005,perets08}. \cite{mer06} studied the mass deficits created by SMBH inspirals evolving in spherical 
symmetric nuclei and found an average $\alpha \sim 0.5$. However, unless their mass $M_{\bullet} \lesssim {\rm few} \times 10^6 M_\odot$, 
the evolution of SMBH binaries in gas-poor {\it spherical} galaxies tends to stall due to depletion, on the (local) dynamical timescale, 
of the pool of stars whose orbits intersect the binary \citep[e.g.][] {MM03,MF04,ber05,kh11,Preto11}. The binaries 
studied by \cite{mer06} did not reach coalescence due to the FPP. Therefore they estimated mass deficits
at the time when a hard binary forms in their simulations. \cite{MMS2007} extended the latter study by 
following the inspiral of binaries up to coalescence in the lower mass range using an approximate Fokker-Planck description
for stellar scattering. They find substantially higher mass deficits than before, albeit restricting their calculations to spherical
models and very shallow cusps (with initial $\gamma \sim 1/2$) in strong contrast with the larger $\gamma$ typical of compact 
low-mass nuclei. More precise estimates of $\alpha$ from more realistic galaxy merger studies are therefore certainly needed;
in particular, ones that do not rely on any of the following approximations: spherical symmetry, low $\gamma$, crude  
model for inferring the mass ejected from the increase in the binary's orbital energy, or those generally inherent to the 
Fokker-Planck formalism. In this respect, preliminary studies of mass deficits created by the merger of equal mass galaxies, 
with moderate SMBH mass ratio, moderate central slope $\gamma \sim 1$, and that generate mildly triaxial remnants, suggest 
that the resulting mass deficits could be substantially higher $M_\mathrm{ej} \sim 2 M_{\bullet}$ within $\lesssim 3 r_\mathrm{h}$ 
\citep{Preto09} -- a result which we confirm and generalize in this paper. Note, however, that alternative scenarios for 
explaining the formation of cores have been proposed \citep{gm08}. 

%Numerical studies of SMBH binary evolution in spherical galaxies show stalling of the binary due to depletion of stars that intersect the b%inary orbit \citep[e.g.][]
%{MM03,MF04,ber05,kh11}. So the 
%coalescence can be delayed if there is no mechanism to derive the stars back to the ``loss-cone''.  The torques from gas, if 
%present, \citep[e.g.][]{escala05,dotti07,cuadra09}, or loss-cone repopulation due to ``massive perturbers'' \citep{perets08} can derive the% binary's further evolution. 
%In the absence of gas the binary can enter the domain controlled by gravitational waves only if the galaxy geometry allows for a much large%r population of 
%centrophilic orbits than in the spherical case \citep{MP2004,ber06}. 

In this paper, we revisit these problems by detailed $N$-body simulations using a range of more realistic models of merging
galactic nuclei with varying mass ratio $q$ and central logarithmic slopes $\gamma$. We restrict the simulations to 
$q=M_\mathrm{gal,s}/M_\mathrm{gal,p}=M_{\bullet 2}/M_{\bullet 1}$. We study the shape of the merger remnant of galaxies having different density 
profiles and different mass ratios. For each of our models, the hardening rate of the 
SMBH binary, the axis ratios of the merger remnant and the resulting mass deficit are calculated. In section 2, we describe our 
initial models and numerical methods used in our study. The evolution of SMBH binary in galaxy mergers is described in section 3, 
as well as the time evolving densities and shapes of newly formed galaxies resulting from the galaxy merger. Section 4 describes 
the analysis and results of ``mass-deficits'' resulting from the galaxy mergers. The coalescence timescales for binary SMBHs from 
the time of formation of binary till the full coalescence of SMBHs due to emission of gravitational waves are presented in 
section 5. Finally, section 6 summarizes and discusses the results of the paper.

\section{Initial conditions and numerical methods}\label{sec-model}

\subsubsection{The host galaxies and their SMBHs}

Our aim is to study the dependence of the hardening rates with the mass ratios and central 
concentration of the nucleus -- as well as the resulting imprint of the SMBH inspiral on the stellar
distributions of post-coalescence nucleus.
Following closely the initial set-up of our previous numerical experiments (see papers I and II),
we represent the individual galaxies or galactic nuclei by spherically symmetric $N$-body models. 
The density profile of the models follows the Dehnen (1993) profile of the form
\begin{equation}
\rho_\mathrm{D}(r) = \frac{(3-\gamma)M_\mathrm{gal}}{4\pi}\frac{r_{0}}{r^{\gamma}(r+r_{0})^{4-\gamma}}, \label{denr} \, 
\end{equation}
\noindent
where $M_\mathrm{gal}$ is the total mass of the galaxy, $r_{0}$ is the scale radius, and $\gamma$ is 
the inner logarithmic slope. For $r \gg r_0$, the density declines with a logarithmic slope of $-4$, 
which, although not close to the exterior density of real galaxies, provides a convenient cut-off 
that avoids the waste of CPU time in following stars whose orbits are too far from the
SMBH binary to have any significant impact on its long term evolution. The resulting cumulative mass 
profile is given by
\begin{equation}
M_\mathrm{D}(r) = M_\mathrm{gal}\left(\frac{r}{r+r_{0}}\right)^{3-\gamma}. \label{massr} 
\end{equation}
In addition, we represent the SMBHs by massive particles with zero velocity
placed at the center of both the primary and the secondary galaxies. The masses of the SMBHs are set
to be $0.5$ percent of their host galaxy. This value is few times higher than the currently accepted value 
of $M_{\mathrm{gal}}/M_{\bullet}$ $\sim$ $~0.001$ \citep{fer00,geb00,mer01,har04}. However, some galaxies may have larger value of this ratio than others. 

%With this choice we are cutting out the mass in the
%exterior of our galaxy models when compared to the real ones.  }
%, a choice made to be consistent with the observed ratio 
%of SMBH mass to bulge mass \citep{kr95,FF05}. 
Thus the two SMBHs in the simulations have the
same mass ratio $q$ as their host galaxies. The Dehnen models have an isotropic distribution function
of velocities (no net rotation), are spherically symmetric and are initially set-up to be in dynamical
equilibrium with a gravitational potential given by $\Phi(r) = -GM_\bullet/r+\Phi_*(r)$, where $\Phi_*(r)$
is the gravitational potential due to the stars alone. 

The galaxies size ratio scales with the corresponding mass ratio as $R_2/R_1 \propto \sqrt{M_2/M_1}$ . 
By choosing four different values of $\gamma = 0.5, 1.0, 1.5, 1.75$, the central part of our galaxy 
models represent the variety of observed density profiles in both early and late type galaxies.

For Dehnen models the ratio of the effective radius (projected half-mass radius) $R_\mathrm{e}$ to the
half-mass radius $r_{1/2}$ is given approximately as 
\begin{equation}
\frac{R_{e}}{r_{1/2}} \approx 0.75, \label{Reff} 
\end{equation}
\noindent
and shows only a weak dependence on $\gamma$. 
In the following we refer to the more massive galaxy as the {\em primary} galaxy and the lighter one as the 
{\em secondary}. In our model units we use for our primary galaxy a total mass and scale radius of $M_\mathrm{gal,p}
 = r_\mathrm{0,p} = 1$. We also set the gravitational constant to $G=1$.

In the absence of relativistic effects on the binary's orbit, our models are scale-free and can be {\it a posteriori}
scaled to galaxies of different total mass and size. However, the inclusion of radiation reaction effects on
the binary due to GW emission introduces an absolute scale associated to the universal value of the speed of
light. Such scalings are described in detail in section 5 when coalescence times are explicitly computed.

The effective radii of our models can be compared to the effective radii in observed galaxies to calculate 
$r_{0}$. By fixing galaxy mass $M_\mathrm{gal}$ our models can be scaled to physical units using following 
relations:
\begin{mathletters}
\begin{eqnarray}
\left[T\right] &=& \left(\frac{GM_\mathrm{gal,p}}{r_{0,p}^{3}}\right)^{-1/2}\nonumber\\
    &=& 1.5 \mathrm{Myr} \left(\frac{M_\mathrm{gal,p}}{10^{11}M_{\sun}}\right)^{-1/2} \left(\frac{r_{0,p}}{\mathrm{kpc}}\right)^{3/2},\\
\left[V\right] &=& \left(\frac{GM_\mathrm{gal,p}}{r_{0,p}}\right)^{1/2}\nonumber\\
    &=& 655 \mathrm{km\,s^{-1}} \left(\frac{M_\mathrm{gal,p}}{10^{11}M_{\sun}}\right)^{1/2} \left(\frac{r_{0,p}}{\mathrm{kpc}}\right)^{-1/2} \ .
\end{eqnarray}
\end{mathletters}

We have performed a series of $N$-body experiments where we vary the mass ratio of the galaxies and SMBHs $q$ 
and the inner density slope $\gamma$.  In Table~\ref{TableS} we summarize the model parameters and properties 
for the primary galaxies used in our simulations. The computational wallclock time increases with increasing 
cuspiness (higher $\gamma$) of the density profile since particles in the center need very 
small numerical timesteps in order to resolve accurately the (locally) strongly varying gradients of the
gravitational potential. In order to keep the computational time within reasonable limits we built our model D 
(highest $\gamma=7/4$) with only half as many particles as the other models. 

\begin{table}
\caption{Model parameter and properties of the primary galaxy.}
\begin{center}
\begin{tabular}{c c c c c c c c }
\hline
Model & $N$ & $\gamma$ & $r_{1/2}$ & $R_{\mathrm{eff}}$\\
\hline
A & 128k & $0.5$ & $3.13$& $2.35$\\
B & 128k & $1.0$ & $2.41$& $1.81$\\
C & 128k & $1.5$ & $1.69$& $1.27$\\
D &  64k & $1.75$& $1.35$& $1.01$\\
\hline
\end{tabular}\label{TableS}
\tablecomments{Columns
 from left to right show the model name, the particle number $N$,
 the half-mass radius $r_{1/2}$ and the effective radius $R_{\mathrm{eff}}$,
 respectively, for our primary galaxy models.}
\end{center}
\end{table}

 The secondary galaxies used in our merger simulations have the same density profile,
 i.e., the same $\gamma$, as the primary galaxies, but have different masses $M_{\mathrm{gal,s}}$
 and scale radii $r_\mathrm{0,s}$. Both primary and secondary galaxy models have equal number of particles N.
%{\bf THIS MEANS WE HAVE A TWO MASS COMPONENT STELLAR SYSTEM, WITH MASS RATIO INCREASING
%WITH q, AM I RIGHT? IF SO, THERE IS AN ``UNCONTROLLED'' FLUX OF MASSIVE PARTICLES FALLING
%DUE TO DYNAMICAL FRICTION --- ESPECIALLY IN THE q=0.1 CASES... }
%A summary of the main parameter for all merger simulations as
%well as the mass ratio $q$ of the galaxies is given in Table~\ref{TableA}.
 %This value is roughly $5$ times higher than the usual $M_{\mathrm{gal}}/M_{\bullet}$ $\sim$ $~0.001$. 
%With this choice we are cutting out the mass in the
%exterior of our galaxy models when compared to the real ones.  
  
% \subsubsection{Galaxy merger setup}
The initial center of mass positions and velocities for the two galaxies
are calculated from the Keplerian orbit of the equivalent two-body problem, with given 
apo- and peri-centers,
$r_\mathrm{a}$ and $r_\mathrm{p}$, respectively. The two galaxies start at the apo-center of 
their relative orbit. The initial separation between the center of mass of 
the two galaxy is $\Delta r = 15 r_\mathrm{0,p}$; and since the half-mass radius of each 
nucleus is $R_{1/2} \lesssim 2.5$, this choice ensures that the galaxies are
initially well separated while, at the same time, we minimize computing time.
The initial relative velocity of the two galaxies is chosen such that the 
SMBH separation at first peri-center passage is $\sim 2r_\mathrm{0,p}$; in other words, 
the initial orbit of the binary galaxy has eccentricity $\sim 0.75$, which 
corresponds to a circularity parameter value of $\epsilon=L/L_\mathrm{c} \sim 0.66$ in
rough agreement with the values measured from full galaxy scale merger simulations
\citep{Kazan10}.

%For a typical, luminous ($M_\mathrm{B}\approx-20$)
%elliptical galaxy or bulge, $r_\mathrm{e}\approx 1.5$\, kpc, 
%$M_\mathrm{gal}\approx 10^{11}M_{\sun}$ and $[T]\approx 1.1$\,Myr.
 %Different sets of simulations are carried out using different value of $\gamma = 0.5, 1.0, 1.5$ and $1.75$.
 %Each of this set has four value of $q = 0.1, 0.25, 0.5, 1.0$. For $\gamma = 1.0$, we also run a merger
 %of galaxies with $ q = 0.05$. Tables \ref{TableA} summarizes the parameters of the galaxy merger models.
  
\begin{table}
\caption{Parameters of the galaxy mergers study} 
\centering
\begin{tabular}{c c c c c c c c c}
\hline
Run & $N_\mathrm{tot}$ & $\gamma$ & $q$ & $r_\mathrm{0,s}$\\
\hline
A1 & 256k & $0.5$& $0.1$& $0.316$\\
A2 & 256k & $0.5$& $0.25$& $0.5$\\
A3 & 256k & $0.5$& $0.5$& $0.707$\\
A4 & 256k & $0.5$& $1.0$& $1.0$\\[0.2ex]
B1 & 256k &	$1.0$& $0.1$& $0.316$\\
B2 & 256k &	$1.0$& $0.25$& $0.5$\\
B3 & 256k & $1.0$& $0.5$& $0.707$\\
B4 & 256k & $1.0$& $1.0$& $1.0$\\
B5 & 256k & $1.0$& $0.05$& $0.22$\\[0.2ex]
C1 & 256k & $1.5$& $0.1$& $0.316$\\
C2 & 256k & $1.5$& $0.25$& $0.5$\\
C3 & 256k & $1.5$& $0.5$& $0.707$\\
C4 & 256k & $1.5$& $1.0$& $1.0$\\[0.2ex]
D1 & 128k & $1.75$& $0.1$& $0.316$\\
D2 & 128k & $1.75$& $0.25$& $0.5$\\
D3 & 128k & $1.75$& $0.5$& $0.707$\\
D4 & 128k & $1.75$& $1.0$& $1.0$\\
\hline
\end{tabular}\label{TableA}
\tablecomments{Col.(1) Galaxy merger model. Col(2) Number of particles. Col(3) Central density profile index $\gamma$.
 Col(4) Mass ratio of merging galaxies and their black holes. Col(5) Scale radius of secondary galaxy.}
\end{table}

\subsection{Numerical Methods and Hardware}

The $N$-body integrations are carried out using the $\phi$GRAPE\footnote{\tt
ftp://ftp.ari.uni-heidelberg.de/staff/berczik/phi-GRAPE/} \citep{harfst}, a parallel,
direct-summation $N$-body code that uses general purpose hardware, namely Graphic Processing
Units (GPU) cards to accelerate the computation of pairwise gravitational forces between all
particles.
Our updated version of the { $\phi$GRAPE} (= {\bf P}{\it arallel}
{\bf H}{\it ermite} {\bf I}{\it ntegration} with {\bf GRAPE}) code
uses the 4-th order Hermite integration scheme to advance all
particles with hierarchical individual block time-steps,
together with the parallel usage of GPU cards to calculate the
acceleration ${\vec  a}$ and its first time derivative ${\dot{{\vec a}}}$ 
between all particles. For the force calculation we use 
the {\tt SAPPORO} \citep{gab09} library which emulates on the GPU the standard
GRAPE-6 library calls.

As { $\phi$GRAPE} does not include the
regularization \citep{mikkola98} of close encounters or binaries,
we have to use
a gravitational (Plummer-)softening between all components. The
softening length is chosen to be sufficiently 
small (i.e. $5 \times 10^{-5}$ in our model units) to keep the (dense) stellar
system as collisional as possible/necessary. 
In the few runs in which we include the $\mathcal{PN}$ terms to the equations
of motion of the SMBH binary, the softening between the SMBH particles is set equal to 
zero. Our implementation of the $\mathcal{PN}$ terms is identical to that used in 
\citet{berent09}, except that we cut-off the  $\mathcal{PN}$ expansion at the 
next order. We stress that the binary is, in all cases, integrated taking into 
account the gravitational field from the stellar cluster.

The $N$-body integrations were carried out on three computer clusters. ``{\tt titan}'', at the
Astronomisches Rechen-Institut in Heidelberg, the
GPU-enhanced cluster ``{\tt kolob}'' at the University of Heidelberg
and ``{\tt laohu}'' in Beijing.

For a detailed discussion on the $N$-body methods and choice of 
parameters, we refer the reader to paper I.

\section{SMBH BINARY EVOLUTION IN GALAXY MERGERS}\label{Resutls}

\begin{figure}
\centerline{
  \resizebox{0.95\hsize}{!}{\includegraphics[angle=270]{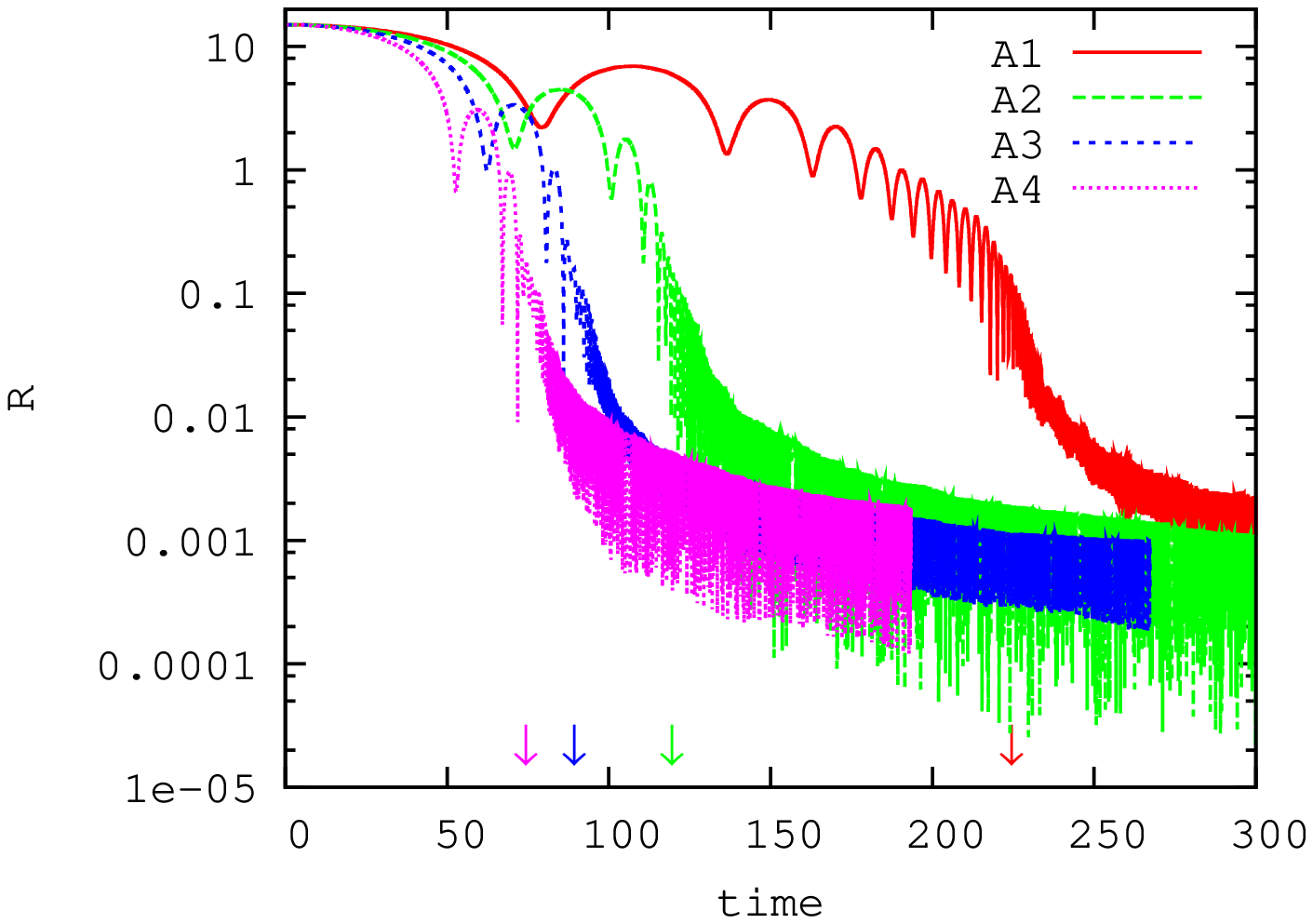}}
  }
\centerline{
  \resizebox{0.95\hsize}{!}{\includegraphics[angle=270]{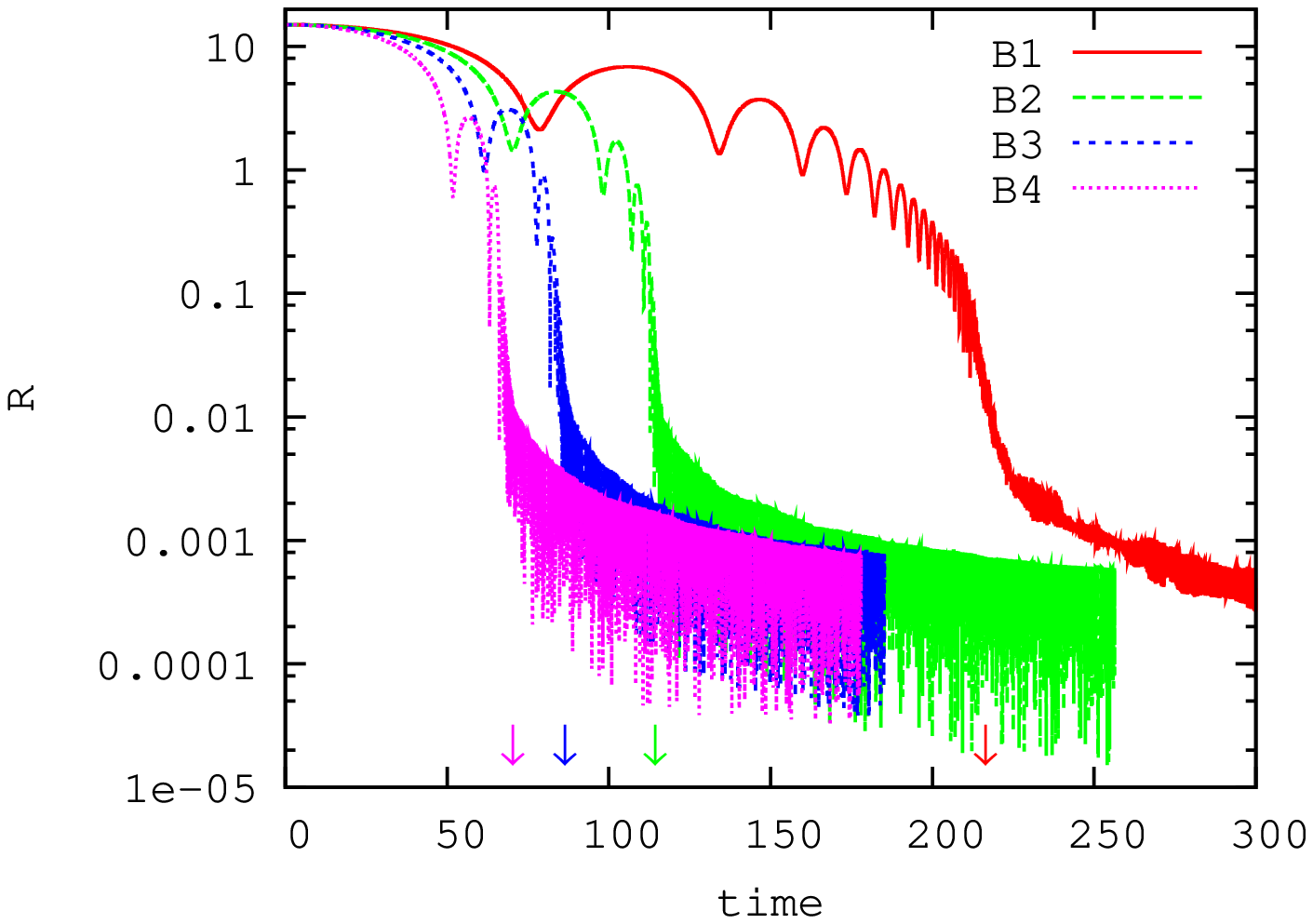}}
  }
  \centerline{
  \resizebox{0.95\hsize}{!}{\includegraphics[angle=270]{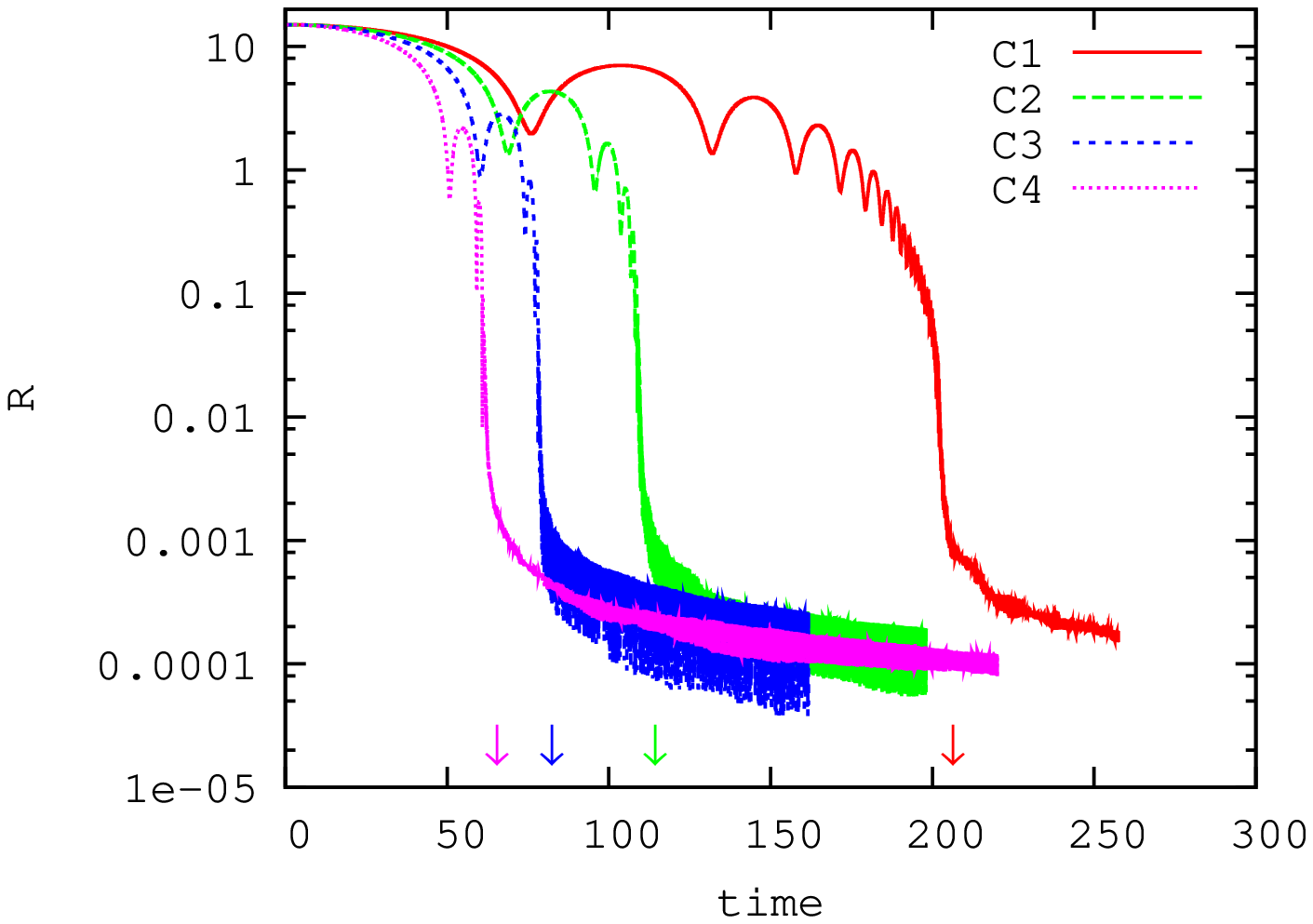}}
  }
  \centerline{
  \resizebox{0.95\hsize}{!}{\includegraphics[angle=270]{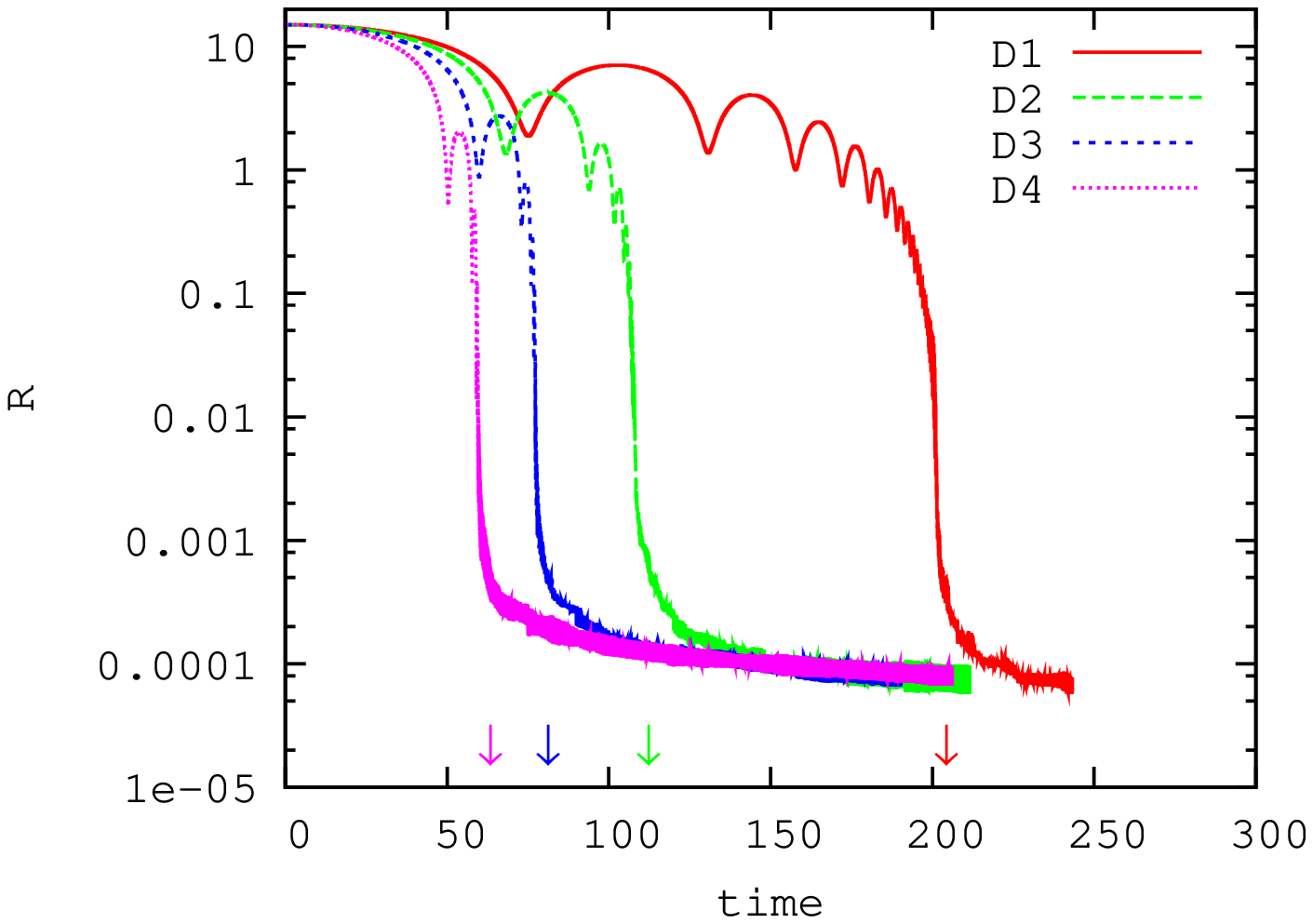}}
  }
\caption[]{
Evolution of the separation between the SMBHs, in $N$-body integrations of galaxy mergers for 
$\gamma = 0.5 ,1.0,1.5$ and $1.75$ (top to bottom) according 
to Table \ref{TableA}. The arrows represent the time ($T = 0$) at which the two black holes become 
gravitationally bound for each model.  Time and $R$ are measured 
in $N$-body units.
} \label{sep}
\end{figure}

Figure \ref{sep} shows the relative separation $R$ between the two black holes during a galaxy merger.
According to the equivalent two-body trajectory used to set up the initial galaxy orbital parameters, 
the galaxy centers, and hence the SMBHs, should 
reach a separation of $2r_\mathrm{0,p}$ during the first pericenter passage. However, as figure \ref{sep} shows, in case
$q \gtrsim 1/2$, the SMBH separation shrinks well below $2r_\mathrm{0,p}$ during the galaxy merger phase. 
%This shrinking of the orbit of the two galaxies -- with
%respect to the corresponding two-body orbit -- is caused by dynamical friction. Dynamical friction is more efficient 
%for the mergers with high $q$ values. 
As $q$ decreases, and hence the mass of the secondary galaxy decreases as well, 
%dynamical friction becomes less efficient; and not only it takes a longer time to bring the SMBHs close enough 
%to form a binary, but
 the galaxies reach the first pericenter passage after following more closely the corresponding two-body 
orbit. In the same figure the color of the arrows signals the time $T$ at which the two SMBHs in a 
particular model form a binary system. As $q$ decreases, the arrows move from left to right signaling the longer 
interval of time between the start of merger and formation of the binary.
Neither the time $T$ nor $R_\mathrm{peri}$ are as sensitive to $\gamma$ as they are to $q$. 

When the SMBHs become bound their separation is $R \approx r_\mathrm{h}$. Once a SMBH binary is formed, dynamical friction 
and the ejection of stars by the gravitational slingshot, act together to efficiently extract angular momentum from 
the massive binary and the binary separation shrinks rapidly.  When the semi-major axis $a \approx a_\mathrm{h}$, this 
initial rapid phase of binary hardening comes to an end as the pool of stars whose orbits cross the binary orbit
gets depleted on the (local) dynamical timescale.\footnote{The semi-major axis $a$ and eccentricity $e$ of 
the binary are defined here via the standard Keplerian relations, i.e., neglecting effects of the field
 stars.} 

The subsequent hardening of the binary occurs only if the loss cone orbits are replenished.\footnote{The loss cone is 
the region of phase space corresponding, roughly speaking, to orbits that cross the binary, i.e. with angular 
momentum $J \lesssim J_\mathrm{lc} = \sqrt{GM_{\bullet}fa_\mathrm{bin}}$, where $f={\mathcal O}(1)$ \citep{1977ApJ...211..244L}.} In a 
spherical galaxy with no gas, the only dynamical mechanism for replenishing loss cone orbits is two-body relaxation, 
with the corresponding timescale for repopulating the loss cone orbits scaling as $\sim N/\ln N$ \citep{BT08}, and 
the inspiral rate thus becomes strongly $N$-dependent \citep{MF04,ber05}. In papers I and II, it was shown that the 
hardening phase, in the case of a mildly triaxial remnant nuclei resulting from a galaxy merger, is essentially 
$N$-independent -- which allows our results to be scaled to real galaxies whose typical number of stars, $N_\mathrm{star}$, is 
always several orders of magnitude larger than the maximum number reachable with state-of-the-art direct $N$-body simulations. 
Furthermore, it was also shown in Papers I \& II that the flux of stars into the loss cone in a triaxial remnant 
is also consistently higher than in the spherical case, meaning the binary evolution and coalescence can occur 
much faster than in spherical models. 
%The departure from spherical symmetry ensures a constant supply to the binary to interact with.

\begin{figure}
\centerline{
  \resizebox{0.95\hsize}{!}{\includegraphics[angle=270]{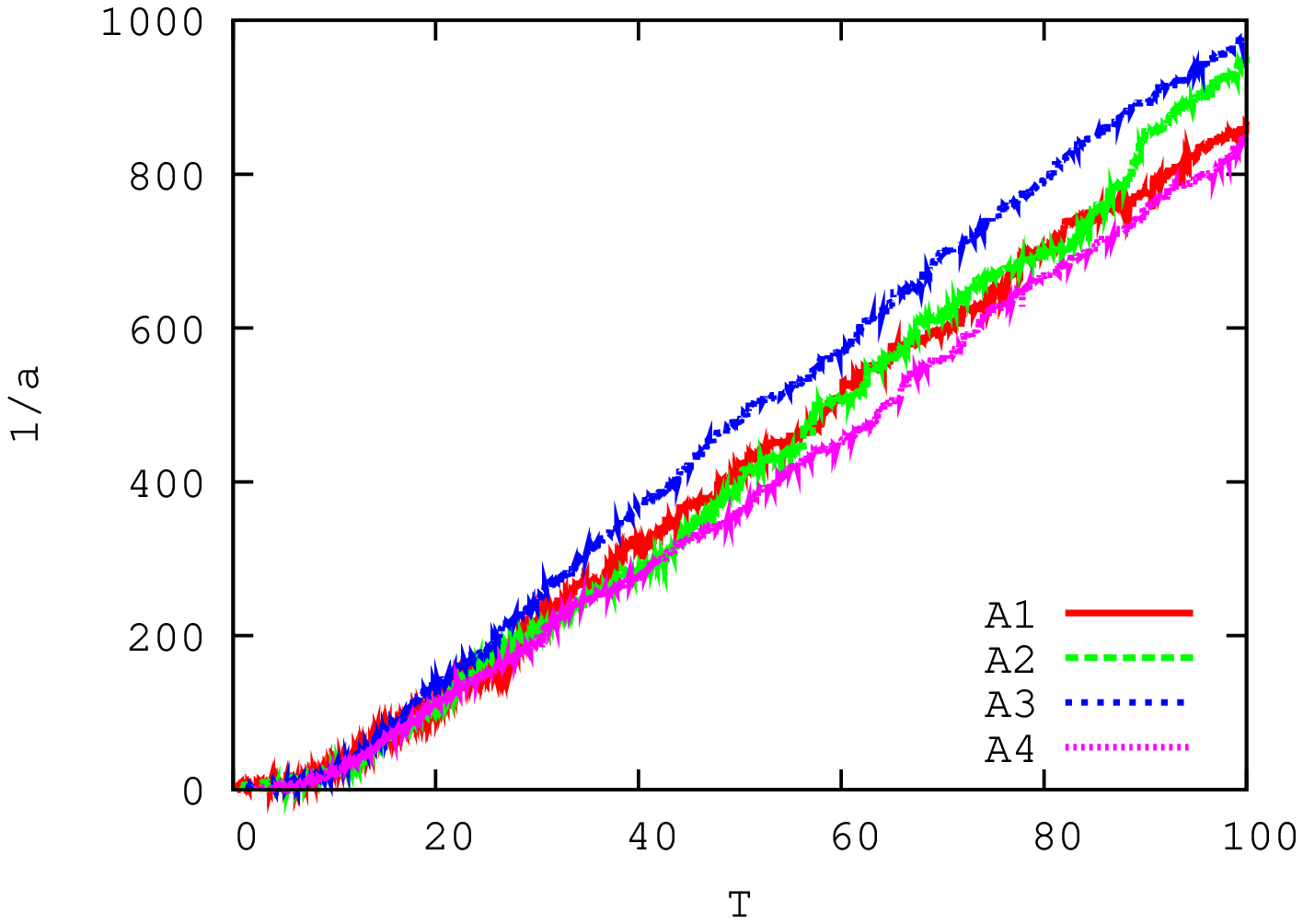}}
  }
\centerline{
  \resizebox{0.95\hsize}{!}{\includegraphics[angle=270]{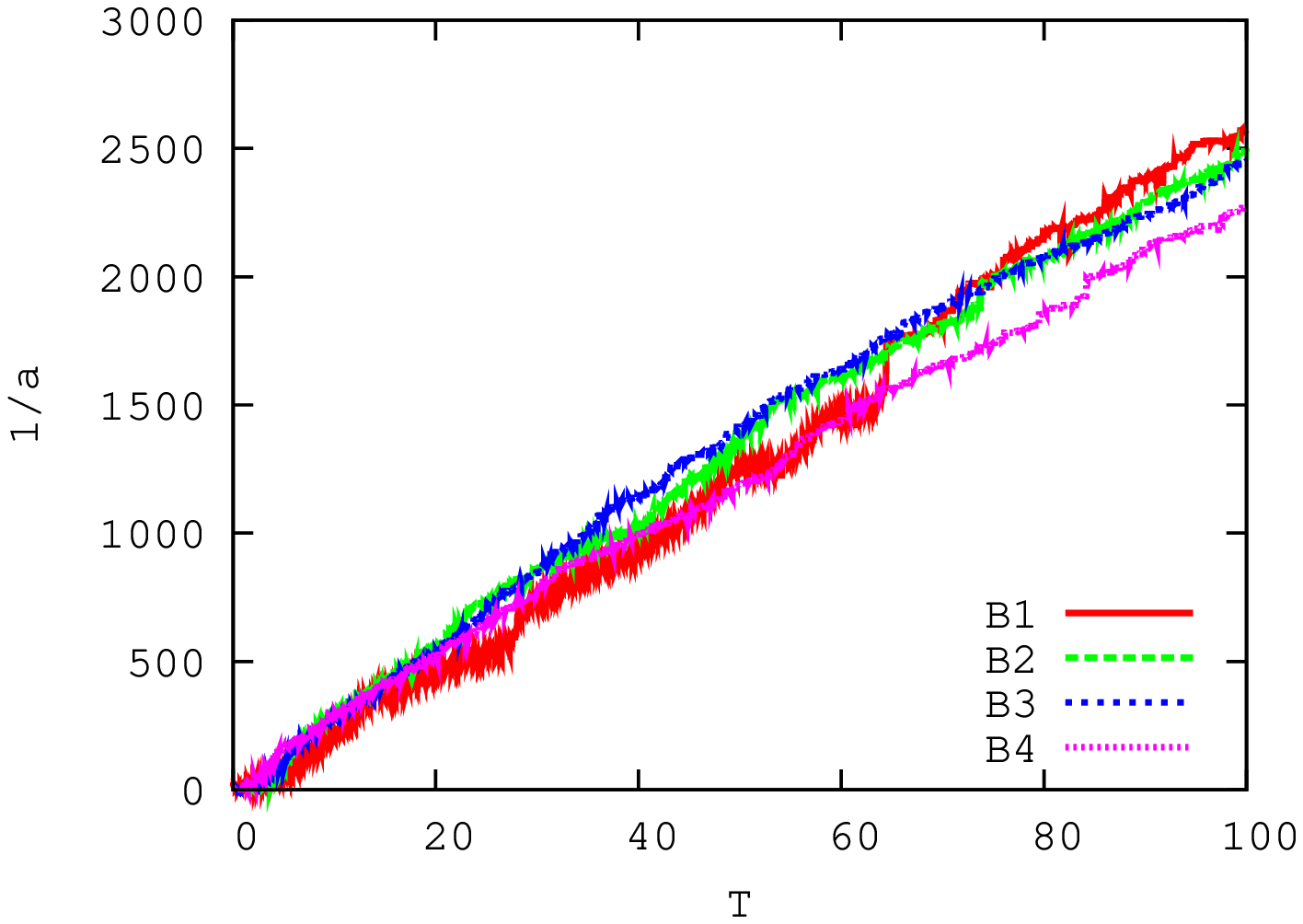}}
  }
  \centerline{
  \resizebox{0.95\hsize}{!}{\includegraphics[angle=270]{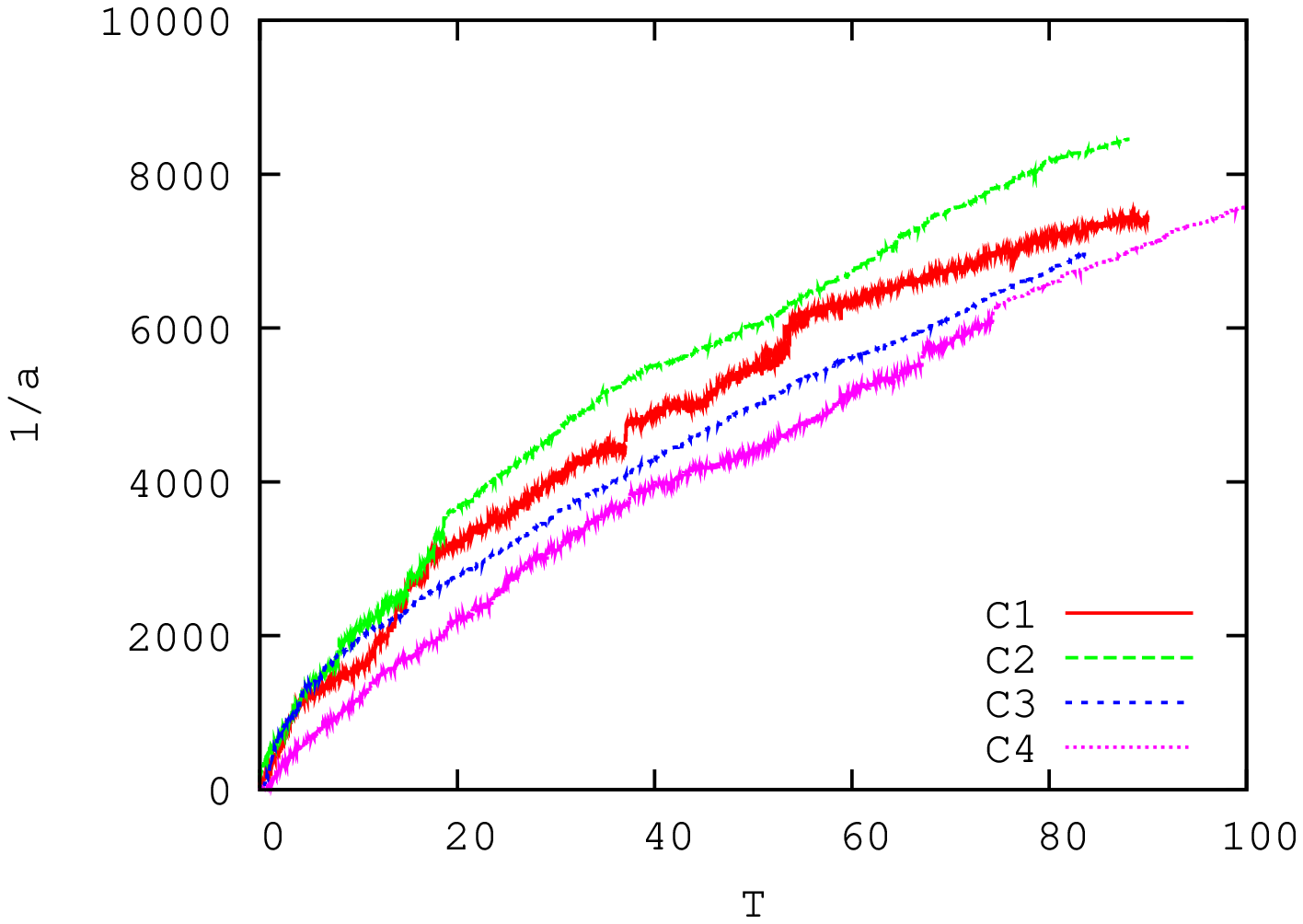}}
  }
  \centerline{
  \resizebox{0.95\hsize}{!}{\includegraphics[angle=270]{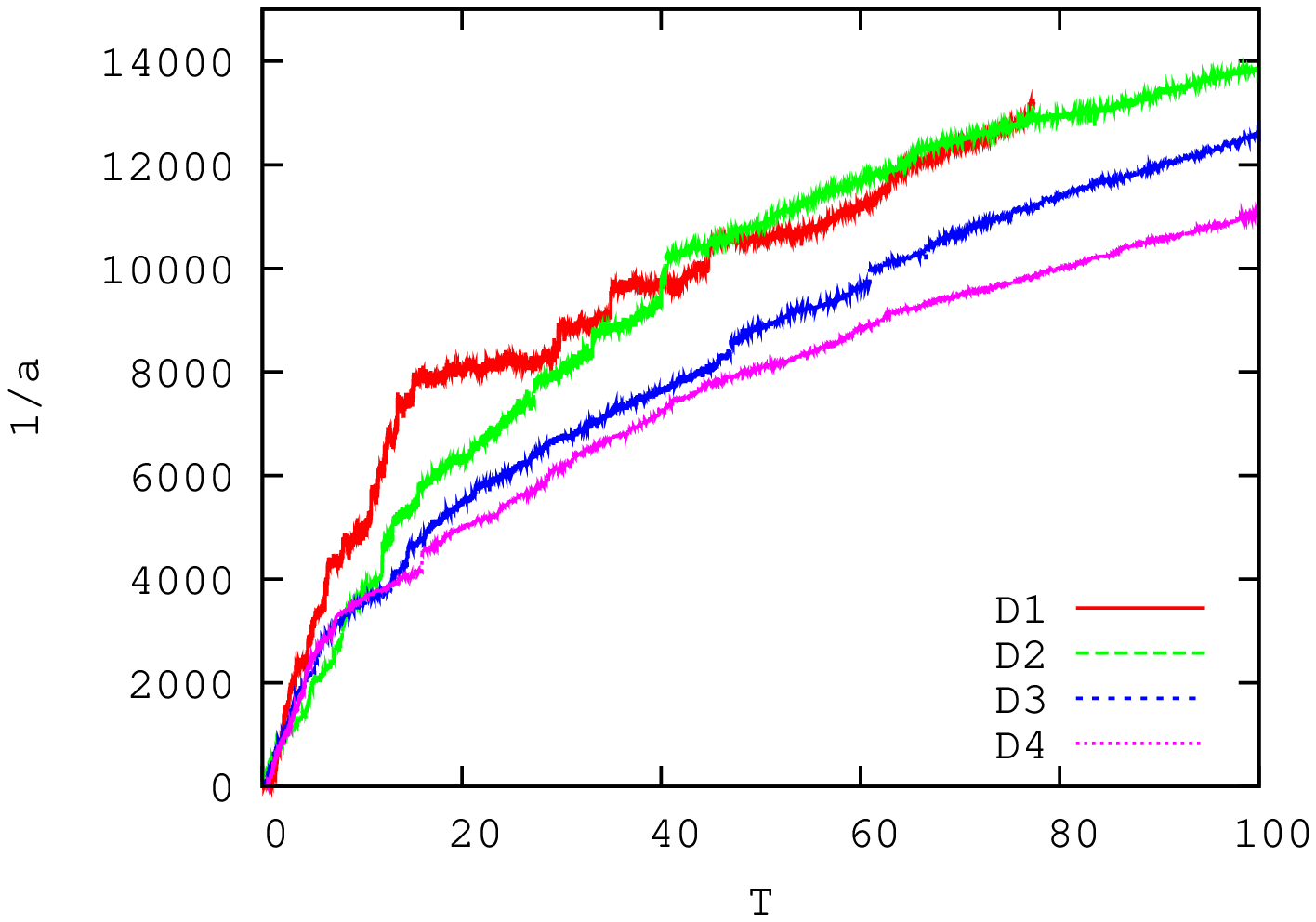}}
  }
\caption[]{
Evolution of SMBH binary semi-major axis, in $N$-body integrations of  galaxy mergers  for 
$\gamma = 0.5, 1.0, 1.5$ and $1.75$ (top to bottom) according to 
Table \ref{TableA}. $T$ measures the time since the instant when the SMBHs formed a bound pair.

} \label{semimajor}
\end{figure}

\begin{figure}
\centerline{
  \resizebox{0.98\hsize}{!}{\includegraphics[angle=270]{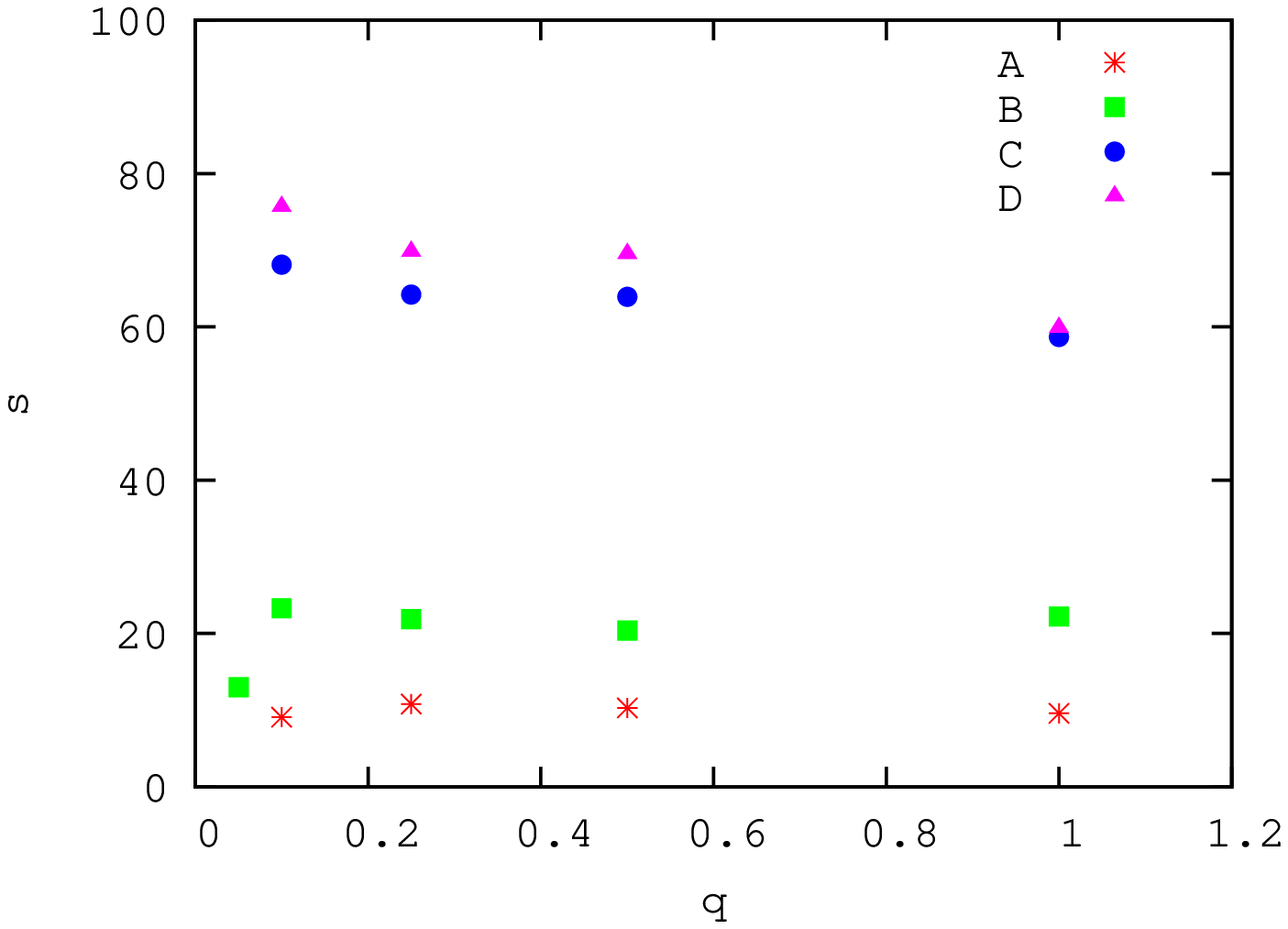}}
  }
\caption[]{
Average hardening rates  for $\gamma = 0.5 ,1.0,1.5$ and $1.75$    
according to Table \ref{TableA}. The average is measured between $T=60$ and $T=100$ in $N$-body
units. See text for details.
} \label{hrates}
\end{figure}

Figure \ref{semimajor} shows the time evolution of the binary's inverse semi-major axis $1/a$ from the time $T$ when 
the two black holes become bound. An initial phase with faster binary hardening becomes more evident for steep inner 
density profiles ($\gamma = 1.5, 1.75$) -- presumably corresponding to the clearing of the original loss cone \citep{Yu02}. 
The binary hardening rates $s$ are calculated by fitting straight lines to $a^{-1}(t)$ in the late phase of binary evolution 
from $T = 60 - 100$. The results are shown in Figure \ref{hrates}. The hardening rates depend very weakly on the mass ratio 
of binary SMBHs, in contrast with the $\sim M_{\bullet}$ dependence found in equal mass merger study (papers I \& II) and, 
for fixed $M_{\bullet}$, on the mass ratio $q$. On the other hand, the value of $s(t)$ increases significantly for higher 
values of $\gamma$. Note that for the single run with the smallest mass ratio in our sample, $q = 0.05$ and $\gamma = 1$, 
the hardening rates are significantly weaker than the remaining $\gamma = 1$ models. 

Following Paper II, the SMBH binary hardening rate, which is a measure for the energy loss of the binary, is given by
\begin{equation}
s(t) \equiv \frac{d}{dt} \left( \frac{1}{a} \right) \approx \frac{2 m_* \langle C \rangle}{M_{\bullet} a}
\int_0^{+\infty} dE \mathcal F_{lc}(E,t),
        \label{st}
\end{equation}
where $\mathcal F_\mathrm{lc}(E,t)$ is stellar flux into the loss cone, $\langle C \rangle = O(1)$ is a dimensionless 
quantity obtained from three-body scattering experiments \citep{Q96} and $E=GM_{\bullet}/r + \Phi_*(r)-1/2 \ v^2$
is the (specific) energy of each star ($E>0$ for bound stars). The behavior of the hardening rates can thus be 
qualitatively understood as follows. The flux $\mathcal F_\mathrm{lc}(E,t)$, and its time evolution, depends on the degree
of symmetry in the underlying gravitational potential (or, equivalently, the orbit families supported by it) 
 -- including the radial density distribution of stars around the binary. On the other hand, the flux of stars
into the loss cone is expected to peak around 
$r_\mathrm{h}$ \citep{perets08}. For a given energy $E$ and fixed $M_{\bullet}$, $\mathcal F_\mathrm{lc}(E,t) \propto n(E,t)/\tau(E,t)$ 
where $n(E,t)$ is the number of stars of energy $E$ per unit (specific) energy, while $\tau(E,t)$ is equal, in the 
spherical case, to $\tau_{\mathrm{rlx}} \propto \sigma^3/\rho \propto r^{\gamma-3/2}$ \citep{BT08} or, in the triaxial case, to 
the star's orbital precession time $\tau_\mathrm{prec}(E,t) \propto M_*^{-1/2}(<r) \propto r^{(3-\gamma)/2}$, for $r \lesssim 
{\rm few} \times r_\mathrm{h}$ \citep{Yu02}. Note that, for typical nucleus parameters, $\tau_\mathrm{prec} \ll \tau_{\mathrm{rlx}}$. As a result, 
and since for more concentrated nuclei (higher $\gamma$) $n(E_\mathrm{h},t)$ is higher and $\tau(E,t)$ is shorter, so more
concentrated nuclei experience higher $\mathcal F_\mathrm{lc}$ and higher $s(t)$. 

\begin{figure}
\centerline{
  \resizebox{0.98\hsize}{!}{\includegraphics[angle=270]{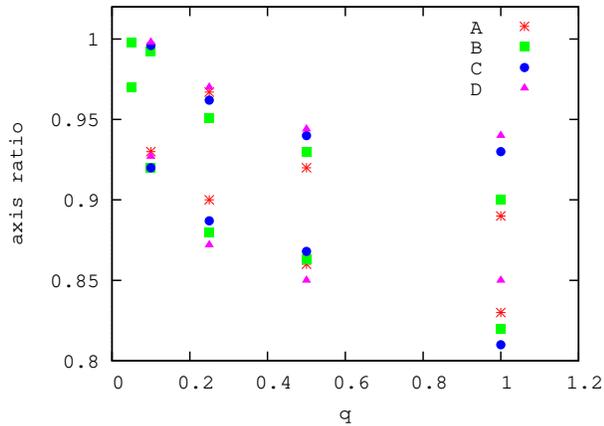}}
  }
\caption[]{
Ratio of intermediate to major (b/a, upper series of points) and minor to major (c/a, lower series of points) axes  for $\gamma = 0.5 ,1.0,1.5 and 1.75$   according to Table \ref{TableA}. 
} \label{triax}
\end{figure}

The triaxiality also plays an important role in determining the hardening rate. This is because of its role in
supporting centrophilic orbits; in fact, at fixed $n(E,t)$ only a fraction of the stars will be on centrophilic
orbits, and this fraction is an increasing function of increasing triaxiality. As $q$ decreases, the triaxiality
of the remnant also decreases as shown in Figure~\ref{triax}. This results presumably in a weaker 
hardening rate since the fraction of stars on centrophilic orbits is smaller. This effect partially 
explains why the dependence of the hardening rate on $M_{\bullet}$ is much weaker than expected based on the 
results obtained in Papers I and II for equal-mass mergers.

We analyze the shape of the merger remnant by calculating principal axis ratios and density profiles for 
galaxy merger models. Figure~\ref{triax} shows the principle axis ratios of the merger remnant; these were 
defined as the axis ratios of a homogeneous ellipsoid with the same inertia tensor. The departures from 
spherical symmetry become more apparent for mergers with larger $q$. For $q = 0.05 \ {\rm and} \ 0.1$, the
merger-induced triaxiality becomes residual and the remnant results only a slightly flattened system. The 
dynamical effect of this transition seems to be abrupt, as the hardening rates shown in Figure~\ref{hrates} 
are essentially independent of mass ratio down to $q \sim 0.1$. In such minor mergers, 
the secondary galaxy gets tidally disrupted within a few pericenter passages, and eventually the {\it naked} 
black hole of satellite spirals in due to dynamical friction. The evolution of SMBH binary in such a merger 
results very similar to those of binaries embedded in spherical galaxy models. Nevertheless in reality situations 
such as these are not very likely to happen as both galaxies would surely retain some triaxiality or flattening 
from a previous major merger. 

\begin{figure}
\centerline{
  \resizebox{0.95\hsize}{!}{\includegraphics[angle=270]{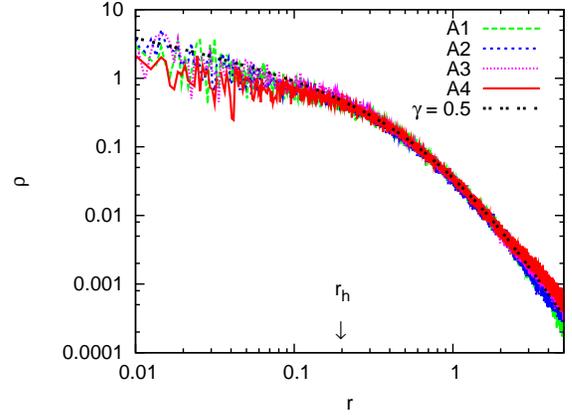}}
  }
\centerline{
  \resizebox{0.95\hsize}{!}{\includegraphics[angle=270]{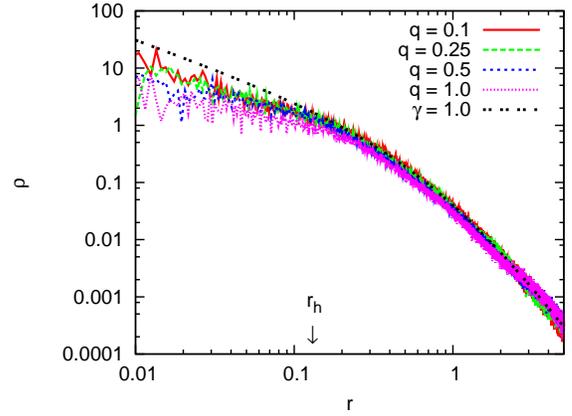}}
  }
  \centerline{
  \resizebox{0.95\hsize}{!}{\includegraphics[angle=270]{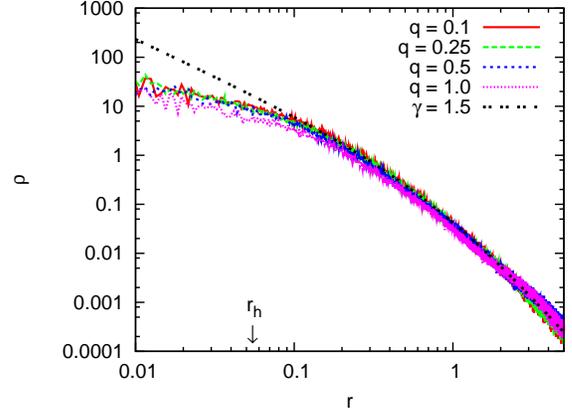}}
  }
  \centerline{
  \resizebox{0.95\hsize}{!}{\includegraphics[angle=270]{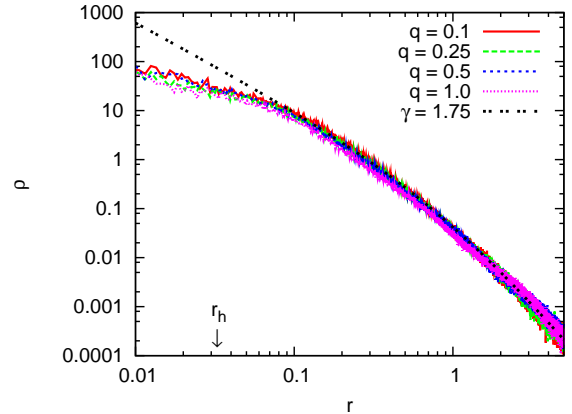}}
  }
\caption[]{
Density profiles of merger remnant  for $\gamma = 0.5 ,1.0,1.5$ and $1.75$ (top to bottom) according to Table \ref{TableA}.
} \label{den-pro}
\end{figure}
The density profiles of the newly formed galaxies as the result of the merger are shown in figure \ref{den-pro}. 
As a result of the galaxy merger and the ejection of stars from the central galaxy region by the inspiraling 
black holes, the merger remnant nuclei have significantly shallower inner slopes than those of its progenitors. 
The "damage" caused by in-falling black holes increases with their mass.  The values of $\gamma_\mathrm{f}$ are calculated 
by fitting Dehnen's model to merger remnant right after the time when phase of rapid binary hardening ends. 
Table \ref{TableC} presents the final inner slopes for merged galaxies. 
%It is interesting to mention that 
%none of our galaxy merger models show completely flat cores ($\gamma < 0.5$) observed at the centers of bright 
%elliptical galaxies.  

\section{Mass Deficits}

Observations of nearby galaxies show that S{\'e}rsic functions provide remarkably accurate fits for the major
axis brightness surface profiles over the main bodies of bulges and early-type galaxies. This has been confirmed
with increased accuracy and dynamic range over the last decade or so \citep{kormendy09}. Moreover, detailed
state-of-the-art simulations of merging galaxies also generate profiles which are consistent with S{\'e}rsic 
functions \citep{Hopkins09p1, Hopkins09p2}. Even though there is not a complete theoretical understanding for 
the remarkable regularity of these profiles, their apparent generality and robustness led people to identify
and interpret departures from S{\'e}rsic fits and use them as a probe of the physics underlying the co-evolution of
galaxies and their massive black holes. 

Cores tend to be found in giant ellipticals and are loosely defined as the central region in a bulge or early-type 
galaxy where the surface brightness deviates and it is below the values that would result from the extrapolation 
of the S{\'e}rsic profile from the main body of the object down to its innermost region. Typically they are 
associated to shallower cusps with $\gamma \lesssim 0.5-1$. One possible explanation of a central core in a
gas poor galaxy can be attributed to the ejection of stars by a hard SMBH binary. The destruction caused by 
the inspiraling massive black hole in the center of galaxy can be measured in-terms of missing mass called 
``mass-deficit". \citet{Gra04} and \citet{Ferrarese06} measured the mass-deficits from the difference in flux 
between the observed galaxies light profiles at the center and the inward extrapolation from their outer S{\'e}rsic 
profiles. Estimated mass-deficits from these studies yield $M_\mathrm{def} \sim (1-2) M_\bullet$, and values as high as 
$\sim (4-5) M_\bullet$ were uncommon. More recent studies obtained larger values with an average value 
$\langle M_\mathrm{def}/M_\bullet \rangle \sim 11$ and an error in the mean of about $18\%$ \citep{kormendy09}.

\begin{figure}
\centerline{
  \resizebox{0.9\hsize}{!}{\includegraphics[angle=270]{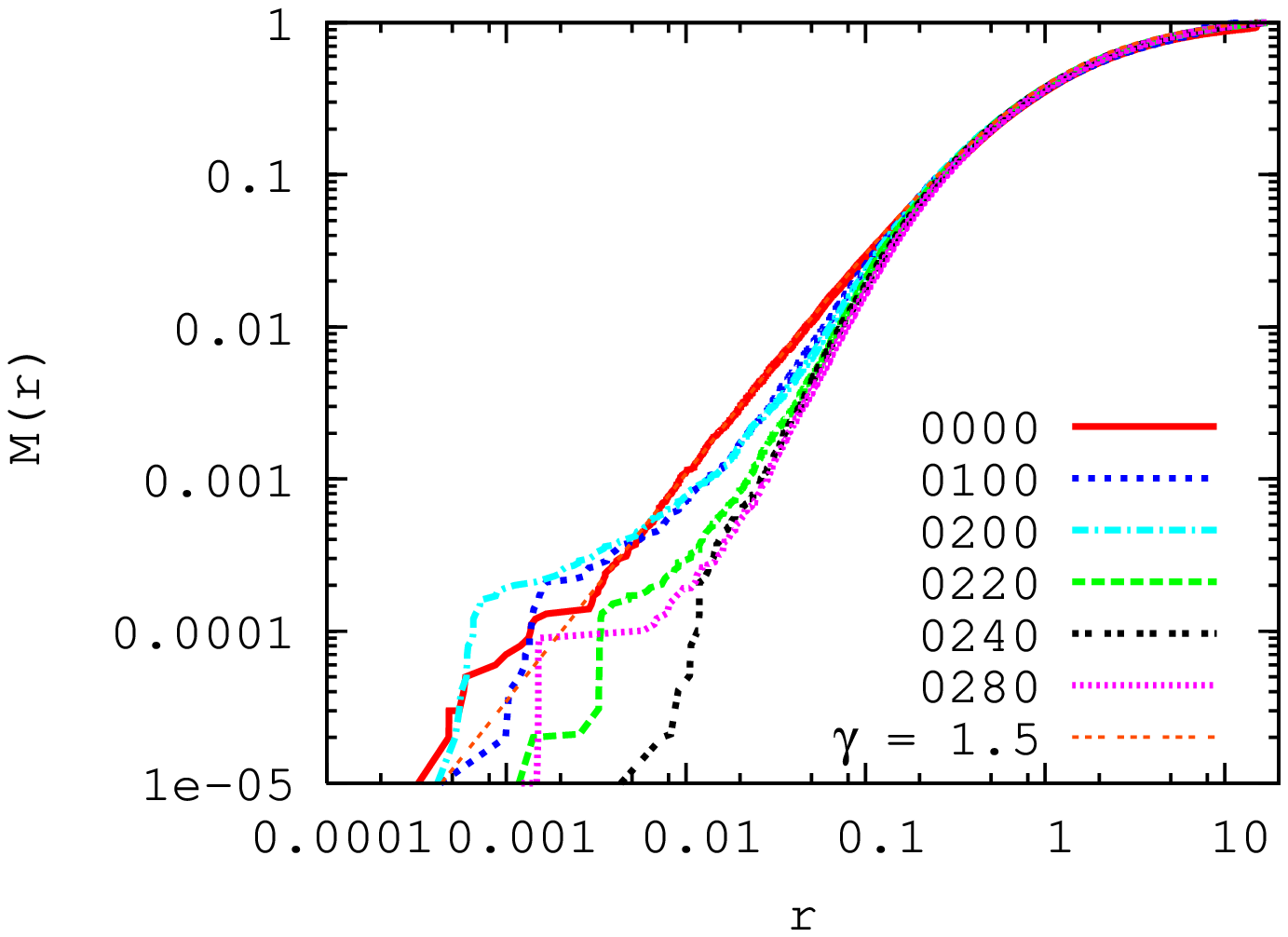}}
  }
\centerline{
  \resizebox{0.9\hsize}{!}{\includegraphics[angle=270]{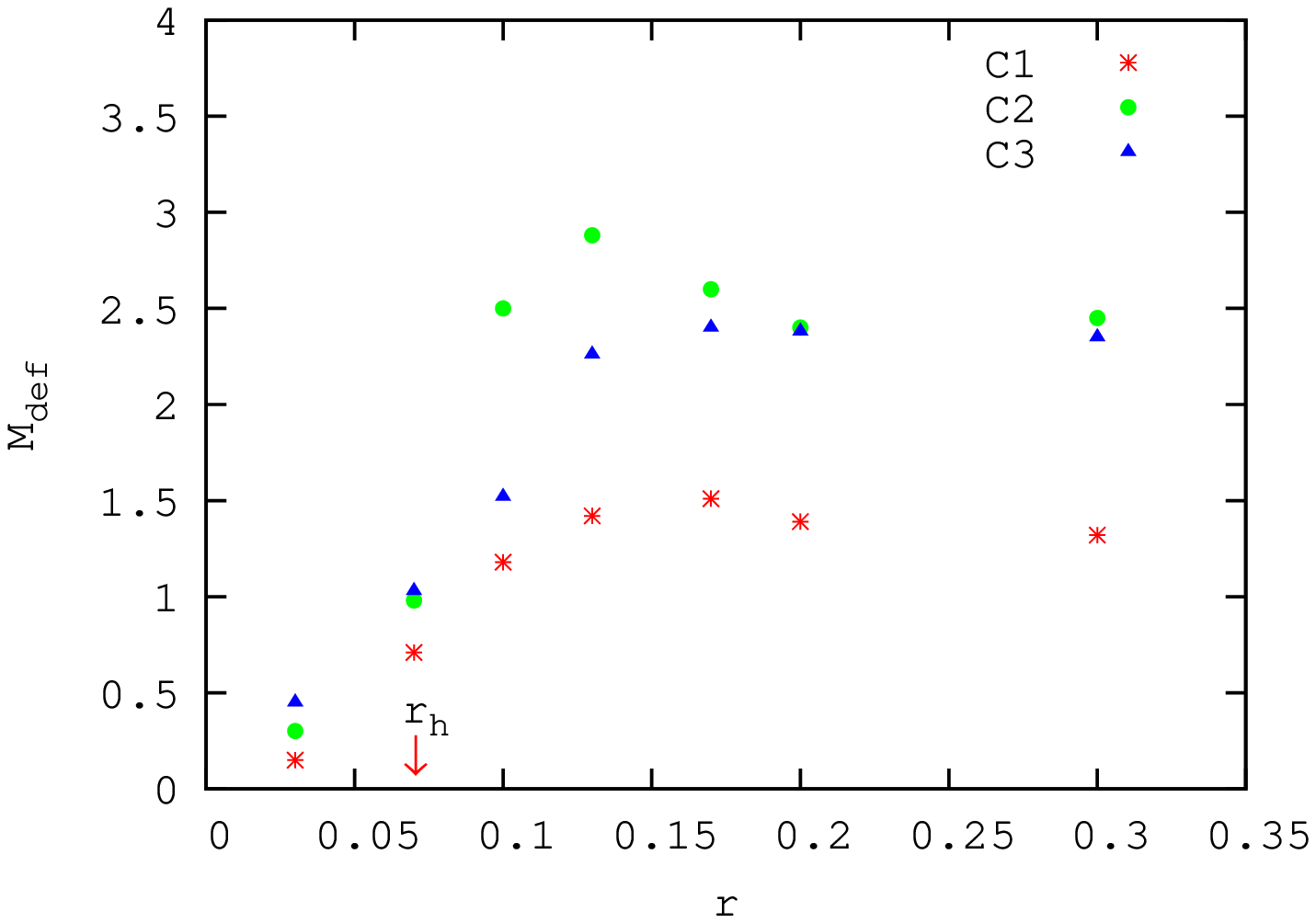}}
  }
\caption[]{
Cumulative mass profile for one of our merger models at different time steps (top) and 
mass deficits normalized to $M_{\bullet}$ for models C (bottom). Central density profiles
at the beginning ($t=0$) and end ($t=280$) are $\gamma=1.5$ and $\gamma \sim 1$, respectively. 
The red arrow in the bottom panel of the figure shows the influence radius at the time of 
the formation of the SMBH binary. 
} \label{mass-pro}
\end{figure}

Top panel of figure~\ref{mass-pro} shows the cumulative mass profile for model C1 at different time of its evolution.
The merger of a secondary galaxy with the primary causes some changes in the inner profile of the latter. This can be
seen by comparing the profiles at times $t=0$ and $t=100$, between which there is a noticeable drop in the stellar
density around $r_\mathrm{h}$ due to the merging of the two galaxies. Then, as the secondary is being tidally disrupted, after 
a few pericenter passages, the profile of the resulting merger remnant remains stable. Once, at time $t \sim 202$, the
two black holes become bound and form a binary, the stars inside the loss cone start to be cleared by the binary and a 
rapid evolution of the mass profile ensues till $t \sim 240$. This drop in the central density is due to the slingshot 
ejection of stars by the binary. As a result, the the mass inside $\sim {\rm few} \times r_\mathrm{h}$ drops rapidly hence changing 
the mass-profile there. Once this phase of rapid mass ejection ends, the density profile -- notably inside $r_\mathrm{h}$ -- remains 
almost constant as the binary continues its inspiral. This can be understood as follows. Inside $r_h$, the gravitational 
potential is approximately spherical and thus the stars residing there can enter the loss cone only by diffusing in energy 
and angular momentum space through two-body encounters. The corresponding timescale is the relaxation time which scales as 
$\tau_{\mathrm{rlx},J} \sim \ \sigma^3/\rho \sim r^{\gamma-3/2}$ \citep{sp87}; since $\gamma \lesssim 3/2$ for all models at almost all 
times of interest, this implies that $\tau_{\mathrm{rlx}}$ increases towards the center. The timescale for such stars to enter 
the loss cone is thus much longer than the typical precession time needed for centrophilic orbits of stars, resident at distances 
$\gtrsim r_\mathrm{h}$, to come sufficiently close to interact with the binary. As a result, the density profiles well inside $r_\mathrm{h}$ 
remain essentially constant during the remainder of the inspiral so the damage impinged on the cusp by the SMBH inspiral 
following a merger is less severe than naive spherical model scenarios might let expect. The conclusion is that it is not very 
likely that a hole in the stellar distribution will ever be created by such mergers.

\begin{center}
\hspace{-3cm}\begin{table*}
\caption{Mass Deficit Analysis}
\centering
\begin{tabular}{c c c c c c c c c c}
\hline
Run &  $r_\mathrm{h}$ & $\gamma_\mathrm{f}$& $a_\mathrm{final}$ & $T_\mathrm{final}$ & $M_\mathrm{def,1}$ & $M_\mathrm{def,1.5}$ & $M_\mathrm{def,2}$ & $M_\mathrm{def,2.5}$ & $M_\mathrm{def,3}$\\
\hline
A1  & $0.19$ & $0.34$ & $6.4 \times 10^{-4}$& $180$& $0.51$& $0.76$& $0.84$& $0.93$& $0.72$ \\
A2  & $0.19$ & $0.26$ & $5.5 \times 10^{-4}$& $180$& $0.50$& $0.94$& $1.20$& $1.20$& $1.05$  \\
A2($\mathcal{PN}$) & $0.19$ & $0.23$ & $merged$& $160$& $0.48$& $0.91$& $1.11$& $1.14$& $1.01$  \\
A3  & $0.19$  & $0.22$ &$5.9 \times 10^{-4}$& $210$& $0.68$& $1.21$& $1.49$& $1.40$& $1.34$ \\
A4  & $0.22$  & $0.18$& $9.7 \times 10^{-4}$& $220$& $0.58$& $0.76$& $1.05$& $1.50$& $1.86$\\
A4($\mathcal{PN}$) & $0.22$ & $0.15$ & $merged$& $204$& $0.59$& $0.74$& $1.06$& $1.47$& $1.81$\\
B1  & $0.155$ & $0.61$ & $2.9 \times 10^{-4}$& $140$& $0.50$& $1.18$& $1.31$& $1.38$& $1.64$\\
B2  & $0.14$  & $ 0.57$ & $3.0 \times 10^{-4}$& $110$& $0.83$& $1.19$& $1.35$& $1.12$& $1.35$\\
B3  & $0.115$ & $0.51$ & $4.0 \times 10^{-4}$& $115$& $1.33$& $1.73$& $2.0$& $2.13$& $1.87$\\
B3($\mathcal{PN}$) & $0.115$ & $0.45$ & $merged$& $168$& $1.69$& $1.75$& $2.10$& $2.30$& $1.91$\\
B4  & $0.13$  & $0.48$ & $3.9 \times 10^{-4}$& $230$& $0.80$& $2.0$& $2.9$& $3.50$& $3.26$\\
B4($\mathcal{PN}$)  & $0.13$ & $0.43$ & $merged$& $189$& $1.05$& $2.05$& $2.80$& $3.34$& $3.21$\\
C1  & $0.066$ & $0.86$& $1.25 \times 10^{-4}$&$090$& $0.71$& $1.18$& $1.42$& $1.51$& $1.29$\\
C2  & $0.066$ & $0.80$ & $1.2 \times 10^{-4}$& $090$& $0.96$& $2.50$& $2.88$& $2.60$& $2.40$\\
C3  & $0.067$ & $0.77$ & $1.4 \times 10^{-4}$& $080$& $1.03$& $1.52$& $2.26$& $2.40$& $2.26$\\
C4  & $0.069$ & $0.70$ & $1.02 \times 10^{-4}$& $140$& $1.10$& $2.3$& $3.9$& $4.80$& $5.30$\\
D1  & $0.045$ & $1.15$ & $8.3 \times 10^{-5}$&$080$& $0.91$& $1.27$& $1.82$& $2.01$& $1.89$\\
D2  & $0.046$ & $1.06$ & $7.1 \times 10^{-5}$& $100$& $1.15$& $1.85$& $2.56$& $3.04$& $3.28$\\
D3  & $0.046$ & $1.04$ &$7.0 \times 10^{-5}$& $112$& $1.39$& $2.36$& $3.14$& $3.81$& $3.97$\\
D4  & $0.047$ & $1.01$ & $7.9 \times 10^{-5}$& $130$& $1.50$& $2.84$& $3.88$& $4.96$& $5.62$\\
\hline
\end{tabular}\label{TableC}
\tablecomments{In this table the columns from left to right represent: (1) Galaxy merger model, 
 (2) the influence radius, (3)The inner density slope $\gamma_\mathrm{f}$ for merger remnant, (4) the separation of the 
two black holes at the end of the run, (5) time at the end of the run in N-body units, (6--10) mass deficits $M_{\mathrm{def},n}$
in units of the mass of the binary $M_{\bullet}$ measured within $n \times r_\mathrm{h}$~;~$r_\mathrm{h}$ is calculated at the time of the formation of the SMBH binary.}
\end{table*}
\end{center}

We calculate the mass-deficits for each of our galaxy mergers as the difference between the stellar mass enclosed
at a given radius at the time when the binary first becomes bound and at the end of the runs. The end of the run is
{\it a priori} a somewhat arbitrary choice since not all runs end at exactly the same point in the binary's inspiral.
However, when we compare the four cases (A2, A4, B3 and B4) for which we followed the binary up to final coalescence, 
we realize that most of the change in the density profiles happens at a relatively early stage in the evolution of the 
binary, and this is mostly covered in almost all of the runs. The conclusion is that the derived mass deficit is quite 
insensitive to the exact final time of the runs, provided this occurs at a time when the binary's semi-major axis already 
reached a separation which is orders of magnitude shorter than the influence radius $r_\mathrm{h}$.

Ours and the (several) observational definitions of mass deficit are obviously not equivalent, but they are surely
somehow related provided the assumption that the mass is ejected by the binary holds true. We will not attempt to 
dwell into the intricacies of a detailed comparison, but will instead use the mass deficits obtained from our simulations
as qualitative indicators regarding the evolution of the galactic nuclei as they undergo minor/major dry mergers. 

The bottom panel of the Figure \ref{mass-pro} shows the resulting mass-deficits measured out to various radii for model C1. 
Mass-deficits have a maximum value at around $2-3 r_\mathrm{h}$ for each model. We are showing the value of mass-deficit at different 
radii in table \ref{TableC} for all the runs. First, we can see that the mass-deficits are clearly larger for more concentrated
(higher $\gamma$) models. They are also increasing with the mass ratio $q$. Note that what we are directly measuring from 
the runs is the effect of the inspiralling binary on the stellar distribution, {\it not} the total amount of mass ejected 
by the binary. The latter value should, by simple energetic arguments and given a fixed total binary mass, be on average 
the same regardless of the detailed properties of the surrounding nucleus. The ``damage" incurred by the stellar cusps in more
concentrated nuclei, however, is expected to be greater as stars on loss cone orbits at small radii will represent a larger
fraction of the (local) total stellar mass than those more outside. If galaxies undergo a series of mergers as 
suggested by the hierarchical galaxy formation scenario, then the mass deficits should add up because the time required to fill in the 
gap (removal of stars by inspiraling SMBH binary) is essentially of the order of relaxation time, which is several orders of
magnitude greater than a Hubble time for 
bright elliptical galaxies \citep{mer06b}. The follow up merger will redistribute the stars in phase space. The mass deficit following $\mathcal{N}$ dry mergers 
is $\mathcal{N} \cdot \left(M_\mathrm{def}/M_\bullet \right)$ so the estimated mass deficits obtained from our simulations imply that at least 
few ($\mathcal{N} = 2-4$) major mergers are required to create $\langle M_\mathrm{def}/M_\bullet \rangle \sim 11$ reported by  \cite{kormendy09}.
%Major mergers ($q \gtrsim 0.25$) of galaxies with shallow 
%cusps ($\gamma=0.5,1$) lead to mass deficits of order $M_\mathrm{def}/M_\bullet \sim 1 - 3$. Taking our results at face value would 
%imply that at least a few major mergers are required to create the average value $\langle M_\mathrm{def}/M_\bullet \rangle \sim 11$ 
%found in \cite{kormendy09}'s sample. 

It is also worth pointing out that the final density profile shown the top panel of Figure \ref{mass-pro} has a central
density slope $\gamma \sim 1$. This means that a single inspiral is not enough to turn a steep cusp ($\gamma \gtrsim 1.5$)
into a core ($\gamma \lesssim 0.5$). 

\section{COALESCENCE TIMES FOR SMBH BINARIES} \label{sec-tcoal}

 In this section we want to derive an estimate for the coalescence times of 
 the SMBH binaries in our simulations, similar to what has been done in
 \cite{berent09}. In the latter work it has been shown that the predictions
 of coalescence times for SMBH binaries in rotating, triaxial galactic nuclei
 agrees well with the results of post-Newtonian $N$-body simulations, up to
 order 2.5PN. 
 To test the accuracy of our estimates made here, we repeat four of our $N$-body
 simulations of merging galaxies -- this time including the post-Newtonian
 equations of motion for the SMBHs up to order 3.5$\mathcal{PN}$ (see Sec.~\ref{PNsim}).
 
 In Fig.~\ref{semimajor}, one can see the rate of change of the binary's inverse
 semi-major axis during the hard binary phase is approximately independent of
 time, so we can write

\begin{equation}\label{eq:dadtnb}
s_\mathrm{NB} = -\frac{1}{a^2} \left(\frac{da}{dt}\right)_\mathrm{NB}  \approx \mathrm{const.},
\end{equation}
a value which we can measure directly from each run.

At some time in the simulation, the semi-major axis $a$ will have fallen to such
a small value that binary shrinking due to GW emission takes over. At such small 
separations -- depending on mass and eccentricity of the binary -- the gravitational 
wave emission becomes so efficient in extracting the angular momentum and energy from 
the binary that the latter becomes effectively decoupled from the rest of the stellar 
system and the SMBHs are led eventually to coalescence as if they were an isolated
binary.

The orbit-averaged expressions -- including the lowest order 2.5$\mathcal{PN}$ dissipative terms -- for 
the rates of change of a binary's semi-major axis, and eccentricity due to GW emission are
given by \citet{P1964}:
%\begin{mathletters}
\begin{eqnarray}
\langle\frac{da}{dt}\rangle_{GW} &=& -\frac{64}{5}\frac{G^{3}M_{\bullet1}M_{\bullet2}M_{\bullet}}{a^{3}c^{5}(1-e^{2})^{7/2}}\times \nonumber \\
&&\left( 1+\frac{73}{24}e^{2}+\frac{37}{96}e^{4}\right),  \label{dadt}\\
\langle\frac{de}{dt}\rangle_\mathrm{GW}  &=& -\frac{304}{15}e\frac{G^{3}M_{\bullet1}M_{\bullet2}M_{\bullet}}{a^{4}c^{5}(1-e^{2})^{5/2}}
\times\nonumber\\
&&\left( 1+\frac{121}{304}e^{2}\right) .  \label{dedt}
\end{eqnarray}
%\end{mathletters}
 
The corresponding hardening rate due to GW emission alone is thus given as:
\begin{eqnarray}
 s_{GW} & = & \frac{64}{5}\frac{G^{3}M_{\bullet1}M_{\bullet2}M_{\bullet}}{a^{5}c^{5}(1-e^{2})^{7/2}}\times \nonumber \\
&&\left( 1+\frac{73}{24}e^{2}+\frac{37}{96}e^{4}\right),  \label{sGW}
\end{eqnarray}

% We chose this approach, because more self-consistent 
%$N$-body simulations including relativistic effects for the SMBH binary \citep[see][]{berent09}, 
%are computationally expensive. \cite{berent09} have shown that in GWs dominated regime Peters 
%formula accurately predicts the coalescence time of the SMBH binary. 

To calculate the coalescence time in our simulations we divide the evolution into two
distinct regimes (see, e.g., \cite{Preto09}): (1) the classical regime, in which the hardening is driven by 
stellar-dynamical effects and (2) the relativistic regime in which the GW emission
is dominant. We define this latter regime, starting from a time $t_0$ when
\begin{equation}
 s_\mathrm{NB} = s_\mathrm{GW} \ . \label{equal_s}
\end{equation}

The semi-major axis $a_0$ at time $t_0$ can be
calculated from Eq.~\ref{equal_s} by assuming that both $s_\mathrm{NB}$
and eccentricity $e(t)$ remain roughly constant\footnote{Note that if $e(t)$ (slightly) 
increases as suggested by scattering experiment by \cite{Q96} and Paper II then
the full time to coalescence shall be smaller than our estimated$t_\mathrm{coal}$. Therefore, our results here should be interpreted as upper bounds to the
true coalescence times.}
during the stellar dynamical hardening phase, as supported by our simulations.

We find that $a_0$ is typically smaller than $a_\mathrm{final}$, the
semi-major axis at the time $t_\mathrm{final}$ at which our simulations end.
Therefore, $t_0$ usually exceeds $t_\mathrm{final}$ and the time interval
$\Delta t$ between $t_\mathrm{final}$ and $t_0$ can be derived as

 \begin{equation}
\Delta t = s_{NB}^{-1}\left(\frac{1}{a_0} - \frac{1}{a_\mathrm{final}}\right).
 \label{tcoal}
\end{equation}

With this, the full time to coalescence, $t_\mathrm{coal}$, in our (Newtonian) simulations
takes into account (1) the time from when the two SMBHs become gravitationally bound 
($T$) until they enter the GW regime $t_{0}$ and (2) the binary`s lifetime $t_\mathrm{GW}$
until the final coalescence. The lifetime $t_\mathrm{GW}$ of an isolated relativistic binary can be
calculated from Equations~\ref{dadt} and \ref{dedt} for a given semi-major axis $a_0$ and
corresponding eccentricity $e_0$, respectively (see Eq.~5.14 in Peters 1964).

%, i.e. when the binary separation reaches six Schwarschild radii.
%Thus, in total, $t_\mathrm{coal}$, is the time when the two SMBHs become gravitationally 
%bound in our simulation until $t_0$, plus $t_\mathrm{GW}(a_0,e_0)$, where $a_0\equiv a(t=t_0), 
%e_0\equiv e(t=t_0)$.

In order to compute the GW evolution rates one requires to adopt some physical units, e.g., of length
 and mass. In 
paper I, we equated the effective radii of our models to the effective radii of observed galaxies to fix the mass 
($M_\mathrm{gal}$) and the length ($r_0$) units. In Paper II, we equate the influence radii of our models to that of
the Galactic center and then extrapolate it to other galaxy masses by assuming the validity of the 
$M_\bullet-\sigma$ relation and a stellar distribution $\rho(r) \propto r^{-7/4}$ at the center. Here to compute 
the coalescence times, we select three different 
elliptical galaxies (M32, M87, NGC4486A) to scale our models to physical units (Table~\ref{scale}). We use the observed mass of 
the SMBH and its influence radius and compare it to the mass and influence radius of the primary galaxy in our models 
to fix $M_\mathrm{gal,p}$ and $r_\mathrm{0,p}$. Model A's which have a very shallow slope are scaled with M87, model B's with 
NGC4486A and model D's are scaled with M32. The choice of galaxies for our models scaling is consistent with the 
fact that bright ellipticals have very shallow inner slopes and faint ellipticals have steep cusps. The estimated coalscence times 
for SMBH binaries in our simulations obtained through the proceduere described above are presented in Table~\ref{GW}.
We find that the time interval spend in the Newtonian and in the relativistic regime, respectively, are always comparable
in almost all our simulations (see column 7 in Tab.~\ref{GW}). This can be understood as follows: binaries with high
 eccentricities reach the relativistic regimes earlier than with low eccentricity and thus shortening the stellar dynamical phase. 
Also the higher eccentricities mean that the GW emission becomes more efficient which then shortens the relativistic
inspiral phase.
 The black hole 
of M32 with a mass $\sim 3 \times 10^{6} M_\odot$ corresponds to possible sources detectable by LISA. The coalescence 
timescales for low mass black hole binaries are less than a Gyr (see table~\ref{GW}) which suggests that 
prompt coalescence of binary SMBH 
detectable by LISA should be very common at high redshifts. Even at the high mass end our models suggest the coalescence 
timescales $\sim$ Gyr or even less depending on the eccentricity of the binary. These timescales are short enough 
that SMBH binaries in these galaxies should achieve full coalescence before a subsequent galaxy merger occurs. 
The coalescence times for SMBHs at low mass end are significantly ($\sim 10 \times$) shorter than the ones presented 
in paper I and more in line with those obtained in Paper II in which the scaling of lower-mass nuclei was set up 
by matching the total mass in stars to that observed in the inner $\sim 2.5$ pc of the Galactic center. The reason 
for this discrepancy is the weak dependence of effective radii with galaxy luminosity relation adopted there, which 
was optimized for nuclei with $\gtrsim 10^8 M_\odot$ black holes, but would lead to overestimate the influence radius 
of the Galactic center (or similarly compact nuclei) by an order of magnitude. Naturally, the coalescence times at 
high mass end of binary SMBHs are in nice agreement between both studies.
\begin{table*}
\caption{Physical Scaling of our Models} 
\centering
\begin{tabular}{c c c c c c c c c c}
\hline
Model & Galaxy & $M_{\bullet} (M_\odot)$ & $r_\mathrm{h}$(pc) & $T (\mathrm{Myr})$ & $L(\mathrm{kpc})$ & $M (M_{\odot})$ & speed of light(c)\\
\hline
A & M87 & $3.6 \times 10^9 $& $460$& $1.9$& $2.3$& $7.2 \times 10^{11}$& $257$\\
B & N4486A & $1.3 \times 10^7 $& $31$& $1.4$& $0.3$& $2.6 \times 10^9$& $1550$\\
D & M32 & $3.1 \times 10^6 $& $3$& $0.75$& $0.12$& $6.2 \times 10^8$& $2011$\\
\hline
\end{tabular}\label{scale}
\tablecomments{Columns from left to right; (1) Primary galaxy model, (2) Observed galaxy used to scale our model, 
(3) Observed mass of SMBH in the galaxy in column 2, (4) observed influence radius, (5)time unit, (6) length unit, 
(7) mass unit and (8) speed of light in model unit.}
\end{table*}

\begin{table*}
\caption{Time to Gravitational Wave Coalescence} 
\centering
\begin{tabular}{cccccccc }
\hline
Run & $a_\mathrm{final}$  & $s_\mathrm{final}$ & $e_0$& $a_\mathrm{0}$ (pc) & $t_\mathrm{0}$ (Gyr) & $t_\mathrm{0}/t_\mathrm{GW}$ & $t_\mathrm{coal}$ (Gyr) \\
\hline
A1 & $6.4\times10^{-4}$ & $9.10$  & $0.50$& $3.5\times10^{-1}$ & $1.30$ & $2.1$ & $1.89$   \\
A2 & $5.5\times10^{-4}$ & $10.8$  & $0.98$& $3.9\times10^{0}$  & $0.12$ & $1.1$ & $0.23$    \\
A3 & $5.9\times10^{-4}$ & $10.3$  & $0.70$& $6.9\times10^{-1}$ & $0.63$ & $1.2$ & $1.15$    \\
A4 & $9.7\times10^{-4}$ & $9.60$  & $0.88$& $1.6\times10^{0}$  & $0.30$ & $1.1$ & $0.57$    \\

B1 & $2.9\times10^{-4}$ & $23.3$  & $0.62$& $7.4\times10^{-3}$ & $2.10$ & $1.3$ & $3.70$   \\
B2 & $3.0\times10^{-4}$ & $21.9$  & $0.98$& $7.1\times10^{-2}$ & $0.24$ & $0.5$ & $0.77$    \\
B3 & $4.0\times10^{-4}$ & $20.4$  & $0.95$& $4.5\times10^{-2}$ & $0.39$ & $1.4$ & $0.66$ \\
B4 & $3.9\times10^{-4}$ & $22.2$  & $0.96$& $6.3\times10^{-2}$ & $0.27$ & $0.7$ & $0.64$  \\

D1 & $9.2\times10^{-5}$ & $75.7$  & $0.69$& $2.1\times10^{-3}$ & $0.48$ & $1.0$ & $0.98$   \\
D2 & $7.1\times10^{-5}$ & $69.8$  & $0.66$& $2.4\times10^{-3}$ & $0.43$ & $1.0$ & $0.82$    \\
D3 & $7.6\times10^{-5}$ & $69.5$  & $0.61$& $2.6\times10^{-3}$ & $0.41$ & $1.4$ & $0.70$    \\
D4 & $7.5\times10^{-5}$ & $59.9$  & $0.60$& $3.3\times10^{-3}$ & $0.38$ & $1.3$ & $0.67$    \\

\hline
\end{tabular}\label{GW}
\tablecomments{Columns from left to right; (1) Galaxy mergers model, (2) semi-major-axis (in model units) at the end of 
simulation $t_{\mathrm{final}}$, (3) hardening rate (model units), (4) eccentricity at $t_{\mathrm{final}}$, (5) semi-major axis of 
the binary at which the stellar dynamical hardening becomes equal to the hardening due to GWs, (6) Life time of binary in 
classical stellar dynamical hardening phase, (7) ratio between the time spend in classical regime and the time spend in GW
regime and (8) Full time to coalescence of SMBHs.  }
\end{table*}

\begin{figure}
\centerline{
  \resizebox{0.98\hsize}{!}{\includegraphics[angle=270]{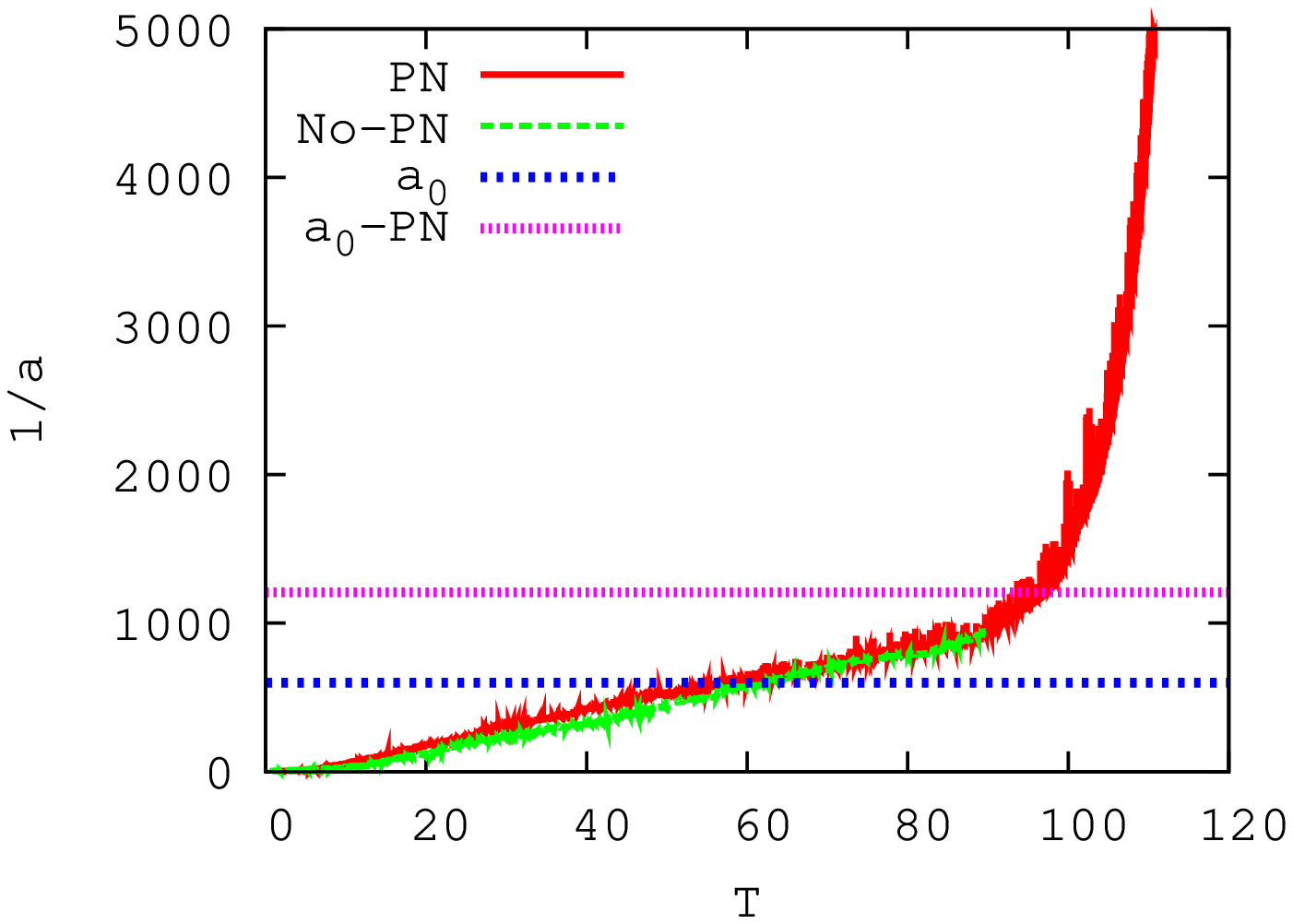}}
  }
\centerline{
  \resizebox{0.98\hsize}{!}{\includegraphics[angle=270]{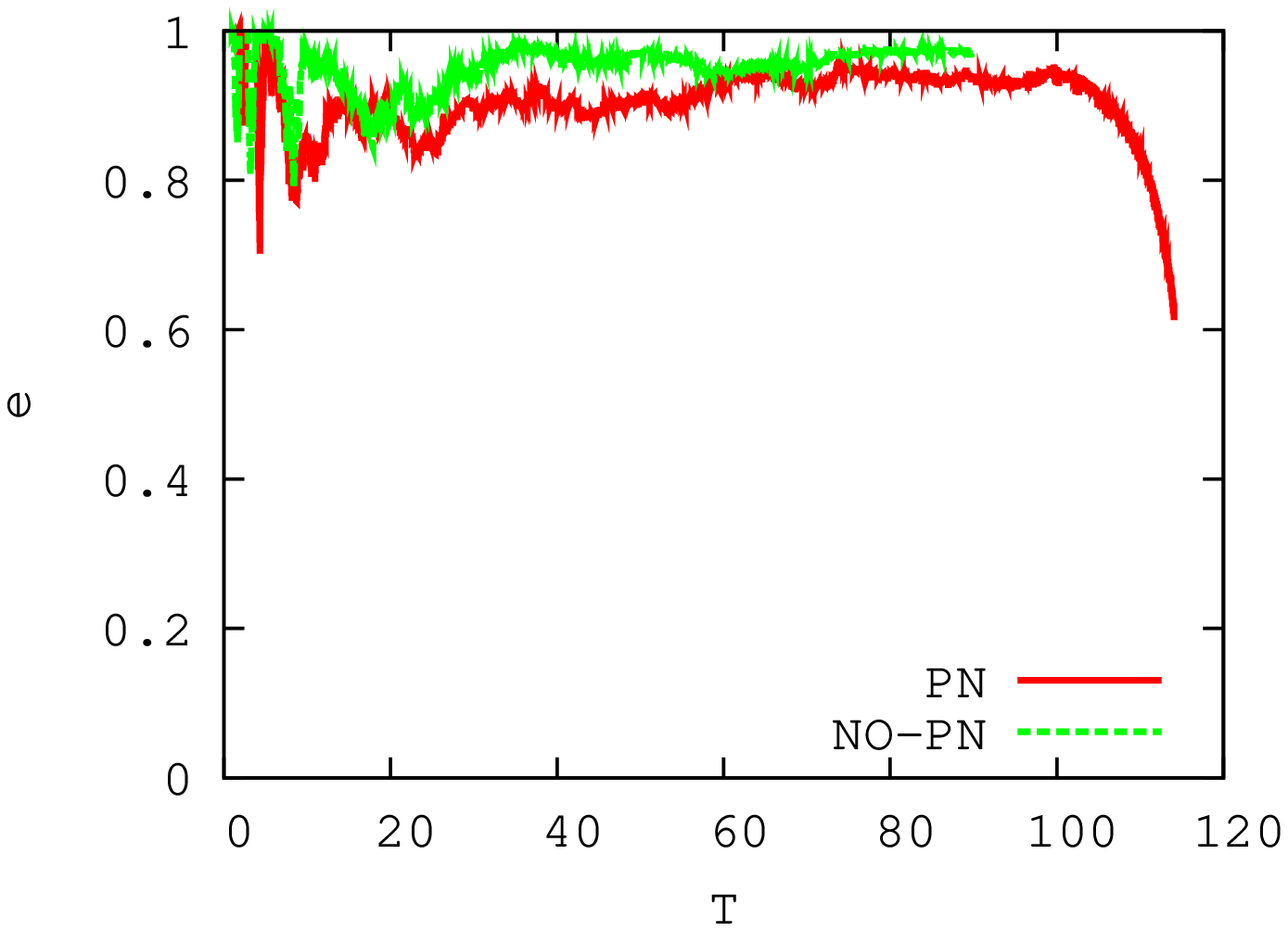}}
  }
\caption[]{
Evolution of the inverse semi-major axis $1/a$ (top) and eccentricity $e$ (bottom) for model A2, with and without 
$\mathcal{PN}$ terms. The horizontal lines represent the estimated semi-major axis of the SMBH binary for which
the stellar dynamical hardening becomes equal to the $\mathcal{PN}$ hardening derived from the run without $\mathcal{PN}$
($a_0$) and with $\mathcal{PN}$ ($a_0-\mathcal{PN}$). See main text for further details.
} \label{figA2}
\end{figure}

\subsection{Post-Newtonian Simulations} \label{PNsim}

In order to verify the accuracy of our estimate for the coalescence times given in Table~\ref{GW}, 
we select the two cases A2 and A4 from Table~\ref{GW} (plus two other cases, $B3$ and $B4$, not shown) 
and re-started these runs with exactly same initial conditions. The choice of these runs is 
motivated by their relative short coalescence times compared to other runs. 

We have implemented the relativistic effects to the MBH binary only by using the 
$\mathcal{PN}$ equations of motion written in the inertial frame of binary center of mass 
including all the terms up to 3.5$\mathcal{PN}$ order

\begin{equation}
\frac{d {\bf v}}{dt} = -\frac{G M_{\bullet}}{r^2} \left[(1+\cal A) {\bf n_{12}} + \cal B {\bf 
v_{12}} \right] + {\cal O} (1/c^8),
\end{equation}

where ${\bf n}={\bf r}/r$, the coefficients ${\cal A}$ and ${\cal B}$ are complicated 
expressions of the binary's relative separation and velocity \citep{B06}. The 
Post-Newtonian approximation is a power series expansion in $1/c$: the $0^{th}$ order 
term corresponds to the dominant Newtonian acceleration. The 1$\mathcal{PN}$, 
2$\mathcal{PN}$ and 3$\mathcal{PN}$ order terms are conservative and proportional to 
$c^{-2}$, $c^{-4}$ and $c^{-6}$. The dissipative 2.5$\mathcal{PN}$ and 3.5$\mathcal{PN}$ 
terms, which are proportional to $c^{-5}$ and $c^{-7}$, cause the loss of orbital energy 
and of angular momentum due to the gravitational wave radiation reaction. We treat the 
SMBHs as point particles and thus we neglect any spin-orbit or spin-spin coupling which 
in general is taken into account in the 1.5$\mathcal{PN}$ (spin-orbit), 
2$\mathcal{PN}$ (spin-spin) and 2.5$\mathcal{PN}$ (spin-orbit) terms in our common 
$\mathcal{PN}$ implementation.

\begin{figure}
\centerline{
  \resizebox{0.98\hsize}{!}{\includegraphics[angle=270]{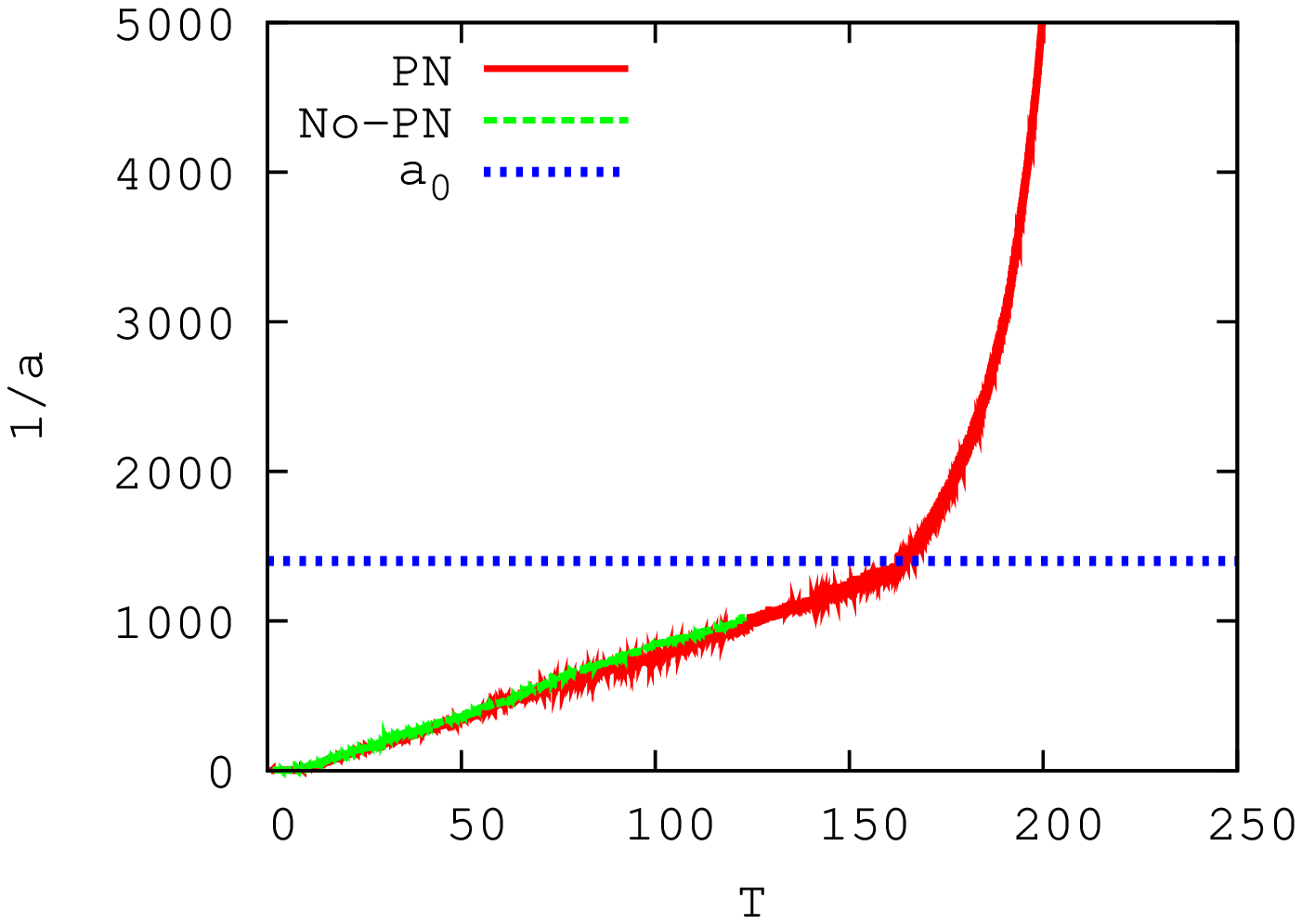}}
  }
\centerline{
  \resizebox{0.98\hsize}{!}{\includegraphics[angle=270]{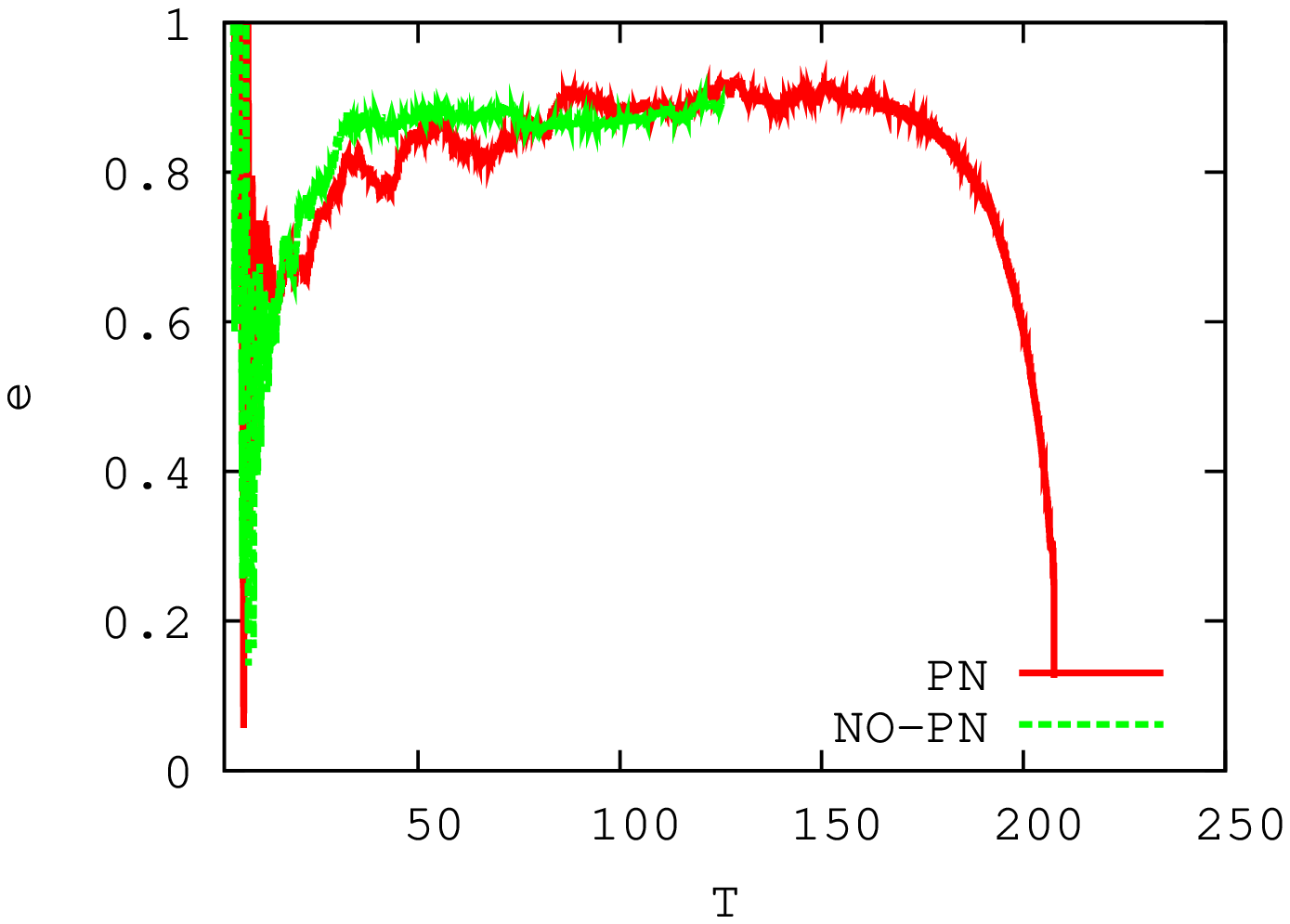}}
  }
\caption[]{
Evolution of inverse semi-major axis $1/a$ (top) and eccentricity $e$ (bottom) for model A4 with and 
without $\mathcal{PN}$ terms. $a_0$ is the semi-major axis of the SMBH binary for which stellar dynamical 
hardening is equal to $\mathcal{PN}$ hardening.
} \label{figA4}
\end{figure}

Figure~\ref{figA2} shows the evolution of the inverse semi-major axis $1/a$ and eccentricity $e$ of the SMBH 
binary for run A2 with and without $\mathcal{PN}$ corrections. Hardening of the binary due to GWs emission 
starts to dominate at $1/a \sim 1000$ which is slightly larger than the estimated $1/a_{0}$
(all in model units). This is due to the slightly higher value of eccentricity reached during the 
run without $\mathcal{PN}$ which is then used to estimate $a_0$. For the correct value of eccentricity
obtained from run with $\mathcal{PN}$, the estimated $1/a_{0}$ matches accurately with the value where hardening due to GWs starts to dominate.  
%%The binary reaches  estimated $1/a_{0} \sim 1400$ in 0.19 Gyrs after becoming gravitationally %%%bound 
%and estimated time is $t_0 \sim 0.14$
The full coalescence time for the binary in our $\mathcal{PN}$ simulations is $0.25$ Gyr which is very 
close to to estimated $t_\mathrm{coal} \sim 0.23$ Gyr. The inverse semi-major axis and eccentricity evolution for 
model A4 are shown in figure~\ref{figA4}. If we look at the evolution of the inverse semi-major axis for this 
run then we see that hardening due to GW becomes important at estimated $1/a_0$, as the average eccentricity 
evolution matches almost perfectly between the $\mathcal{PN}$ and non-$\mathcal{PN}$ runs.

As in the (isolated) models of \cite{berent09}, we find very good agreement between our estimated coalescence times and those found using full 
$\mathcal{PN}$ terms in our merger runs. We should note that the coalescence times for Models D are all in the
upper range of the coalescence times estimated in Paper II for $10^6 M_\odot$ LISA binaries. This can be
easily understood if we note that in the previous paper, the nuclei were all modeled with $\gamma=1$, 
whereas here we choose $\gamma=7/4$ more in line with the expected properties of compact nuclei. We can
see both from Figure~\ref{sep} and Table~\ref{GW} that the binary eccentricities tend to be lower 
($e \sim 0.6-0.65$) for $\gamma=7/4$, and indeed in Papers I $\&$ II, the binaries reached very often values
$e \gtrsim 0.9$ (in accordance to values obtained in this paper for $\gamma=1$). The dependence on 
eccentricity of the coalescence time under GW emission is $T_\mathrm{coal,GW} \sim (1-e^2)^{7/2}$ could easily
account for a decrease of an order of magnitude or so when the binaries are very eccentric. It will be
accordingly very important to further investigate the dependence of the eccentricity evolution under
different values of $\gamma$ and $q$, and this is the subject of a forthcoming paper \citep{MP11}. 

\section{Summary \& Conclusions}\label{sec-concl}

Direct $N$-body simulation of mergers of spherically symmetric galaxies with different mass ratios were performed 
to investigate the evolution of binary SMBHs from the onset of merger through the stellar hardening phase 
until the eventual relativistic coalescence. The merging galaxies have different initial density profiles 
varying from shallow ($\gamma=1/2, 1$) to steep cusps ($\gamma=3/2, 7/4$) -- thus covering the range of stellar
distribution typically observed in the centers of bulges or early-type galaxies. In Papers I and II, we have shown that 
merger-induced triaxiality could support a purely stellar dynamical solution to the FPP in the case of 
equal-mass mergers. In order to assess the prospects for such a solution in the more general unequal-mass 
case, we measured both the hardening rate and merger-induced triaxiality for mass ratios in the range 
$q \in [0.05,1]$. The merger-induced triaxiality found in Papers I and II for equal-mass mergers is still 
present in unequal ($q<1$) mergers, albeit it becomes weaker as $q$ decreases. Minor mergers with $q 
\lesssim 0.05$ of spherical progenitor galaxies leave the primary almost unperturbed, so the subsequent 
binary evolution follows suit in an almost spherical background nucleus. The classical FPP could well 
show up in such cases. This transition seems to be abrupt, as the measured hardening rates are essentially 
independent of the mass ratio $q$ until it suddenly declines at a value somewhere between $q=0.05$ and $0.1$ -- thus 
indicating that only a very modest triaxiality is needed for driving the binary inspiral at rates 
consistently higher than in a spherical nucleus. It is also not surprising that we find that the hardening rate 
increases substantially for high $\gamma$, as more concentrated nuclei will have bigger amounts of centrophilic orbits. 

The solution to the FPP therefore hinges strongly on the expected triaxiality -- or, more precisely,
 the amount of centrophilic orbits supported by it -- at the center of galaxy 
merger remnants. 
%From this perspective, our models can be taken as {\it worst case scenarios}, since 
%they were derived from the merger of {\it spherically symmetric} progenitor galaxies, which is indeed
%somewhat unlikely as most observed ellipticals are significantly flattened and a good fraction of them
%are at least mildly triaxial \citep{KB96,emsellem07}. 
From this perspective, our models may represent {\it worst case scenarios}, since they were derived from the merger of {\it spherically symmetric} progenitor galaxies, which is indeed somewhat unlikely as most observed ellipticals are significantly flattened and a good 
fraction of them are at least mildly triaxial \citep{KB96,emsellem07}. How the shape of the progenitor galaxies
affects the shape of the merger remnant and the resulting SMBH hardening rates is yet unclear.
Nevertheless it is important to pursue increasingly
realistic stellar dynamical models of galactic nuclei in order to refine our understanding of binary SMBH
evolution tracks and associated timescales. A detailed study of the $N$-independence and merger-induced
triaxiality in unequal-mass mergers -- including non-spherical progenitors resulting from previous merger
events -- is being currently pursued \citep{MP11}. 

We have produced estimates of coalescence times using a simplified prescription
for the late relativistic phase of the inspiral -- {\it i.e.} adopting the Peter's (1964) equations for
the evolution of an isolated binary and ignoring the late {\it stellar-driven} eccentricity evolution.
We assessed the accuracy of these approximations by including Post-Newtonian terms (up to 3.5$\mathcal{PN}$
order) to the equations of motion of the binary in a few representative cases of our sample. At least
in these few selected cases, the agreement is remarkably good, and this is mainly due to the fact that
the eccentricity evolution is almost identical in the $\mathcal{PN}$ and non-$\mathcal{PN}$ runs. One 
should add a little note of caution though. The eccentricity evolution may be a quite sensitive function
of the properties of the surrounding stellar cluster (slope $\gamma$, mass ratio $q$, amount of net
rotation, etc) and on the initial conditions of the binary. Previous studies indeed indicate
that one could expect significant evolution of the eccentricity during the hardening phase, even 
though not always of the same sign \citep{sesana10,Preto11,2011MNRAS.415L..35S}. The coalescence
times resulting from our calculations are consistently shorter than a Hubble time at a given 
redshift (up to $z \sim 2-3$ for all masses and up to $z \sim 6-10$ for those in the LISA mass range). 
Coalescence times for binaries of $\sim 10^6 M_\odot$ are all shorter than $1$ Gyr; for the most massive 
$\sim 10^9 M_\odot$ they range between hundreds of millions of years (highly eccentric ones) to $\sim 
1-2$ Gyr (less eccentric ones). SMBH binaries are therefore very promising sources of GWs for LISA 
both at low and high redshift. The coalescence times obtained here -- especially those from models
D -- may be effectively an upper bound if the mild eccentricity growth observed in these runs is
not generic (we have only four runs). In Paper II, we used models with $\gamma=1$ to compute the 
evolution of binaries in the LISA mass range and we obtained faster coalescences; this was partly 
because the binaries were significantly more eccentric. Therefore the detailed properties of the 
distribution of coalescence times hinges in part on the dependence of the eccentricity evolution on 
the properties of the surrounding cluster. The question of whether eccentricity grows, decays or 
remains approximately constant during the hardening phase remains still uncertain and we are 
currently pursuing it \citep{MP11}. Interestingly, the duration of how long the binaries remain
in the stellar dynamical hardening phase and in the relativistic one are roughly equal by an 
eccentricity regulated process. This means that SMBH binaries on highly eccentric orbits have
less probability to be detected in the classical regime but at the same time are sources of
strong GW signals to be detected. On the other hand, binaries with low eccentricities remain
longer in the Newtonian regime and therefore are more likely to be observed in this phase,
but may only be luminous GW sources in the final phase of coalescence. 

With our results we moved another step towards a consistent stellar dynamical solution to the FPP, and
thus of providing a solid dynamical substantiation to cosmological scenarios where prompt coalescences
are the norm during SMBH-galaxy co-evolution \citep{2007MNRAS.377.1711S,2010A&ARv..18..279V}. If our 
results hold true in real galaxies, the bottleneck to SMBH coalescences -- if any -- is likely to be 
associated to the long timescales needed for SMBHs to become a bound pair in galaxy mergers of 
unequal-mass -- especially in case $q \lesssim 0.1$ and gas fractions are low \citep{calleg11}. These 
results are promising for SMBH binaries being abundant LISA sources at high redshift. 

What about the role of gas in hardening the SMBH binaries at sub-parsec scales? Numerical simulations
of merging disk galaxies with a substantial gas-fraction have shown that gas is driven towards the
central region of the merger remnant where it accumulates. As a consequence, the remnant becomes
more axisymmetric (i.e. losing triaxiality) and the orbital structure changes noticeably as compared
to the purely stellar case (Naab, Jesseit \& Burkert 2006). Thus we can expect that the fraction
of centrophilic orbit is affected in the same manner by the presence of the gas. In which direction
it affects the hardening rate is yet an unresolved issue.
%can go in both direction: centrophilic = chaotic and pyramid orbits (Merritt & Poon). The
%merger destroy box-orbits (Naab et al) but may enhance the fraction of chaotic orbits (unclear to
%which extent).

The gas which is funneled to the center of the remnant may undergo strong starbursts which may
lead to the formation of a steep stellar density cusp -- even before the binary has formed --
as shown by Callegari et al. (2009), which then in turn increases the binary's hardening rate
as shown in this work.

The gas could certainly also assist the inspiral if it is cold enough to settle into a circumbinary thin
disk \citep{2005ApJ...634..921A}, even though the uncertainties associated with its long-term dynamical behavior are potentially more
severe than those related to stellar dynamics. For instance, gaseous disks are susceptible to fragmenting
and forming stars. In order for the disk to be at worst marginally stable against fragmentation at every 
radius, its total mass is constrained to be $M_\mathrm{disk} \sim (0.1-0.2)M_{\bullet}$ \citep{cuadra09,lodato09}. This
is not likely to be enough to drive the binary to coalescence.
%Furthermore, as the binary loses angular momentum to the circumbinary disk, gas is transported outwards 
%and the gas disk edge is removed away from the binary thereby weaking the gravitational torques between 
%them. The long term effectiveness of the gas torques for removing angular momentum away from the binary will 
%thus likely depend on the influx of gas from regions just outside the inner parsec -- which even though
%it is not implausible makes the situation certainly more complex and uncertain to model, and it is maybe 
%even more likely to suffer genuinely from a FPP than the stars. (REFERENCES FOR THIS LAST PART!!!)

The "mass deficits" induced to the centers of galaxies with no gas by inspiraling SMBH binaries are found to be
of order $\sim (1-5) M_{\bullet}$, depending on the slope $\gamma$ of their central stellar distribution and on 
their mass ratio $q$. This is consistent with the picture that cores in giant ellipticals are 
scoured by SMBH binary inspirals during successive merger events -- assuming only that relaxation times
are long enough in those systems for the results from different mergers to be cumulative. For galaxies
in the mass range of the Milky Way the picture is basically distinct. Since the amount of mass depletion
in a single merger event only partially destroys a steep inner cusp, it results that even if a major merger 
were to have happened in the Galactic center at redshift $z=1$ or larger, it would have been enough time
to regrow a mass segregated Bahcall \& Wolf cusp since then \citep{PA10}.

\acknowledgments
We thank Gabor Kupi and Patrick Brem for their help in testing and 
implementing the $\mathcal{PN}$ code subroutine for the binary SMBHs
which we use in our phi-GRAPE+GPU code. We also thank referee for his/her valuable comments.

FK was supported by a grant from the Higher Education Commission (HEC) of 
Pakistan administrated by the Deutscher Akademischer Austauschdienst (DAAD).
MP and RS acknowledge supported by the Deutsches Zentrum f\"ur Luft- und
Raumfahrt (through LISA Germany project).
We acknowledge support by the Chinese Academy of Sciences Visiting
Professorship for Senior International Scientists, grant number
2009S1-5 (The Silk Road Project; R.S. and P.B.).
PB and IB both acknowledge financial support by the Deutsche Forschungsgemeinschaft (DFG) through 
SFB 881 "The Milky Way System" (sub-projects Z2 and A1) at the Ruprecht-Karls-Universit\"at Heidelberg.

PB acknowledges the special support by the NASU under the Main Astronomical Observatory 
GRID/GPU computing
cluster project (``{\tt graffias}'' cluster). PB's studies are also partially supported 
by the program Cosmomicrophysics of NASU. 

IB acknowledges funding through a Frontier Innovation grant of the
University of Heidelberg sponsored by the German Excellence Initiative.

We used following GPU cluster to run our simulations.  ``{\tt laohu}''
at the Center of Information and Computing at National Astronomical Observatories, Chinese
Academy of Sciences, funded by Ministry of Finance of People's Republic of China
under the grant ZDYZ2008-2. ``{\tt kolob}'' which has been funded through a Frontier project supported
by the excellence scheme of the University of Heidelberg and by the DFG. ``{\tt titan}'', funded under the grants I/80 041-043 and I/81 396 of the
Volkswagen Foundation and grants 823.219-439/30 and /36 of the Ministry of Science,
Research and the Arts of Baden-W\"urttemberg, Germany.

\end{document}